\documentclass[5p, twocolumn]{elsarticle}
\usepackage{graphicx}
\journal{NIMA}

\begin{document}

\begin{frontmatter}

\title{Construction and Commissioning of Mid-Infrared SASE FEL at cERL}

\author[addkek]{Yosuke Honda}
\author[addkek]{Masahiro Adachi}
\author[addkek]{Shu Eguchi}
\author[addkek]{Masafumi Fukuda}
\author[addqst]{Ryoichi Hajima}
\author[addkek]{Nao Higashi}
\author[addaist]{Masayuki Kakehata}
\author[addkek]{Ryukou Kato}
\author[addkek]{Takako Miura}
\author[addkek]{Tsukasa Miyajima}
\author[addkek]{Shinya Nagahashi}
\author[addkek]{Norio Nakamura}
\author[addkek]{Kazuyuki Nigorikawa}
\author[addkek]{Takashi Nogami}
\author[addkek]{Takashi Obina}
\author[addkek]{Hidenori Sagehashi}
\author[addkek]{Hiroshi Sakai}
\author[addaist]{Tadatake Sato}
\author[addkek]{Miho Shimada}
\author[addkek]{Tatsuro Shioya}
\author[addkek]{Ryota Takai}
\author[addkek]{Olga Tanaka}
\author[addkek]{Yasunori Tanimoto}
\author[addkek]{Kimichika Tsuchiya}
\author[addkek]{Takashi Uchiyama}
\author[addkek]{Akira Ueda}
\author[addkek]{Masahiro Yamamoto}
\author[addaist]{Hidehiko Yashiro}
\author[addkek]{Demin Zhou}

\address[addkek]{High Energy Accelerator Research Organization (KEK), 
1-1 Oho, Tsukuba, Ibaraki 305-0801, Japan}
\address[addaist]{National Institute of Advanced Industrial Science and Technology (AIST),
Central 2, 1-1-1 Umezono, Tsukuba, 305-8568, Japan}
\address[addqst]{National Institutes for Quantum and Radiological Science and Technology (QST), 
Tokai, Ibaraki 3191106, Japan}

\begin{abstract}
\indent
The mid-infrared range is an important spectrum range
where materials exhibit a characteristic response corresponding to their molecular structure.
A free-electron laser  (FEL) is a promising candidate for a high-power light source with wavelength tunability
to investigate the nonlinear response of materials.
Although the self-amplification spontaneous emission (SASE) scheme
is not usually adopted in the mid-infrared wavelength range,
it may have advantages such as 
layout simplicity, the possibility of producing a single pulse,
and scalability to a short-wavelength facility.
To demonstrate the operation of a mid-infrared SASE FEL system
in an energy recovery linac (ERL) layout,
we constructed an SASE FEL setup in cERL, a test facility of the superconducting linac with the ERL configuration.
Despite the adverse circumstance of space charge effects due to the given boundary condition of the facility,
we successfully established the beam condition at the undulators,
and observed FEL emission at a wavelength of 20~$\mu$m.
The results show that the layout of cERL has the potential for serving as a mid-infrared light source.
\end{abstract}

\end{frontmatter}

\newpage

\section{INTRODUCTION}

Human society depends on various materials,
which are being further developed by game-changing innovations in the production of useful materials.
One of the most fundamental properties of a material
is its response to electromagnetic radiation.
In particular, in the mid-infrared spectrum range, 
which is defined as 3-50~$\mu$m in wavelength,
materials exhibit a characteristic response
corresponding to their molecular structure.
By irradiating a narrow-band mid-infrared radiation
that targets one of the oscillation modes of a material,
we can selectively excite the mode in a thermally non-equilibrium way.
It will be useful for efficiently converting one material to another
or for realizing a new state of material that cannot be achieved by using a conventional method
\cite{morichika, forst, maas,  feltusno, toriumi, thzmattermanipuration}.

For such a selective excitation,
a high-intensity and narrow-band light source with wavelength tunability is necessary.
However, in the mid-infrared range,
which is difficult to cover with a conventional laser system,
a suitable light source has not been established at present.
A widely used light source in the mid-infrared range
has been the CO$_2$ laser \cite{co2laser}.
Although it can produce several 10~kilowatts of power,
the wavelength is fixed at 10.6~$\mu$m.
A new technology that has been developing recently
is the quantum cascade laser (QCL).
It can serve as a compact light source and is expected to be used widely.
It can produce 1~W of power in continuous operation
in the approximately 8~$\mu$m wavelength range \cite{qcl}.
The wavelength can be changed 
by modifying the design of the quantum structure of the chip;
however, it does not offer wavelength tunability in operation.

One of the promising approaches is the free-electron laser (FEL) \cite{maday,firstfel}.
Although it requires an accelerator system
and requires a larger facility compared with a conventional laser-based system,
it can produce a high radiation power originating from the beam power,
and the wavelength can be tuned by the undulator setting or the beam energy.
Several infrared FEL facilities have been operating
\cite{clio2, felix, felbe, feltus, novo, kufel}.
FEL systems used in these facilities 
are the optical-cavity-type FEL,
which uses a multi-bunch electron beam,
and the radiation is amplified in an optical cavity
through interactions with many bunches.
The advantage of the cavity-type FEL is that
it can reach a high power even with a low single pass gain,
and hence the requirement for electron beam quality can be relaxed.
On the other hand,
precise optical cavity tuning is necessary
to match the radiation and electron beam in three dimensions.
Changing the wavelength sometimes requires
machine tuning,
and real-time tunability can be a problem.
The cavity-type FEL emits multi-pulse output.
When one wishes to measure the fast response of material
in a single radiation pulse,
a special technique,
such as a fast switch mirror, must be developed \cite{cavitydump}.

The technologies of the superconducting RF accelerator have been widely developed.
When combined with a high-quality electron source,
which has also been developed recently \cite{500kVgun},
it is possible to realize a high-quality electron beam
at a high repetition rate.
Such an electron beam is suitable for use with an FEL working in a single-pass layout
of the self-amplified spontaneous emission (SASE) scheme
at a high repetition rate.

The mechanism of SASE was first explored in the infrared range
\cite{isir, sunshine, clio, bnl, lanl}.
The SASE FEL has now been widely operating in the X-ray range.
However, it has not usually been applied in the long-wavelength range.
%primarily because the cavity-type FEL has been established in this range.
One of the difficulties of the SASE FEL 
in the long-wavelength range
is the time slippage effect in the long undulator.
The electron beam is delayed relative to the radiation pulse
by one wavelength for each period of the undulator.
This effect becomes important when the radiation wavelength is long.
The overlap between the electron bunch and the radiation pulse,
which is essential for the FEL gain,
is limited by the slippage effect.

Recently,
a sub-femto second X-ray FEL has been required 
for the investigation of fast electron dynamics within atoms and molecules.
In such a regime,
the slippage effect becomes important
even at short wavelengths.
In such a short-pulse X-ray FEL,
it has been recognized that the undulator tapering scheme is effective 
in optimizing the efficiency of interaction 
between the radiation and the electron beam in the time slippage
\cite{saldin,ding,fawley,duris}.
When the relationship between the bunch length, the radiation wavelength, and FEL gain length is scaled,
there are similarities between the sub-femto second X-ray FEL
and a mid-infrared SASE FEL.
The techniques established for a short-pulse X-ray FEL 
may be useful for the mid-infrared case, and vice versa.

The cERL \cite{cerl_construction} at KEK is a test accelerator of an energy-recovery linac (ERL \cite{benzvi}) configuration 
with a cw superconducting linac.
It can serve as a model for a future high-beam power accelerator 
utilizing the ERL scheme.
To demonstrate mid-infrared SASE FEL operation
in the modern design of the accelerator, 
we constructed an experimental setup of the SASE FEL 
in the return loop of the cERL.
The typical layout of the ERL
includes a function for controlling the bunch length
with an off-crest operation of the linac cavity
and dispersion tuning in the arc section
\cite{jlabdemo, HAJIMA, alice, sdalinac}.
We established FEL beam tuning by utilizing this function.
The layout at the cERL is not ideal for the FEL.
Due to the relatively low beam energy and the long beam path,
the bunch suffers strong space charge effects 
through the transport line.
To manage this adverse circumstance  
and increase the FEL output power,
we introduced the undulator tapering technique
assuming that the energy chirp in the bunch 
will be counteracted by the tapering.
Finally, we performed a demonstration experiment
using FEL irradiation with a test sample.
It confirmed the establishment of FEL beam tuning 
and operational stability.

The most important advantage of the ERL is the possibility of 
handling a high average beam power.
Because the beam-to-radiation power conversion efficiency of the FEL is quite low,
especially in the short-wavelength range,
a very large beam power will be necessary 
for a high-average-power FEL,  which is required for extreme ultraviolet (EUV) lithography \cite{euverl}.
An SASE FEL in an ERL scheme is a candidate for serving as
such a light source within a reasonable total electric power and beam dump system.
Demonstration of the SASE FEL, even at a long wavelength,
would be a first step in the development of an EUV lithography facility.

In this paper,
we first describe the initial design of the FEL
followed by its detailed simulation in Sec. \ref{sec:principle}.
Then in Sec. \ref{sec:setup},
we describe the experimental setup of the FEL system at the cERL.
Sec. \ref{sec:experiment}
is about beam tuning and the performance of the FEL.
The results are discussed and summarized in
Sec. \ref{sec:discussion}.
The Sec. \ref{sec:conclusion} concludes the paper.

%=========================
\section{DESIGN OF THE SYSTEM}
\label{sec:principle}
%=========================

%---------------------------------------------
\subsection{Conceptual Design of SASE FEL}
%---------------------------------------------

We began the design of this experiment
assuming the boundary conditions of the existing cERL facility.
The target wavelength of the FEL 
was set to the mid-infrared range.
%, especially
%the longer wavelength range than the CO$_2$ laser.
The targeting central wavelength was set to 20~$\mu$m,
considering the electron beam energy of 17.5~MeV.
%This necessitates
%a relatively short period and a narrow-gap undulator.

Before performing a detailed simulation,
we conducted a simple estimation based on 
a one-dimensional steady-state formula
assuming the following expected beam and undulator parameters:
The bunch charge is $Q=$60~pC.
The transverse RMS beam size is $\sigma=$0.2~mm.
The RMS bunch length is $\sigma_t=$1.0~ps.
The undulator strength is $a_w(=K/\sqrt{2}$)=1.0.
The undulator period is $\lambda_u=$24~mm.
The Pierce parameter $\rho$ then becomes \cite{boni}
\begin{equation}
\rho = \left(
\frac{\bar{\kappa}^2 a_w^2 \lambda_u^2 j_0}{16 \pi \gamma^3 I_0}
\right)^{1/3}
\sim 0.01 \quad ,
\end{equation}
where $\bar{\kappa}(=[JJ])\sim0.86$, 
$I_0=mc^3/e\sim17~$kA,
and $j_0=Q/(2\pi \sigma ^2 \sqrt{\pi}\sigma_t)$.
The FEL gain length becomes
$L_g=\lambda_u/(4\pi \sqrt{3}\rho) = 0.1$~m.
The saturation length is
$L_{sat}=\lambda_u/\rho=2.1$~m.

\begin{figure}[htb]
	\begin{center}
 	 \includegraphics[bb=0 0 600 400, width= 0.8\linewidth]{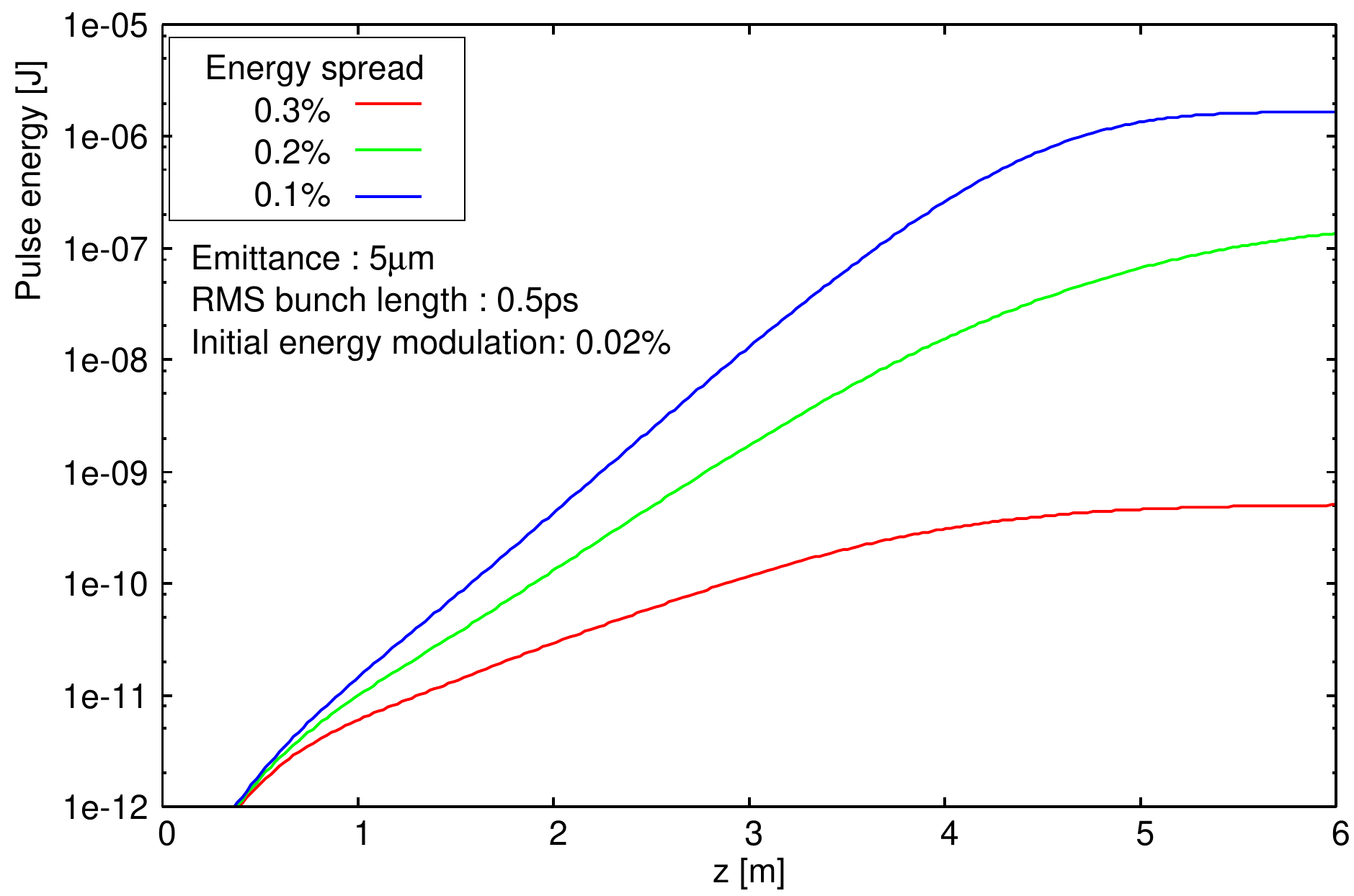}
	  \caption{
	  Simulation for estimating required length of  undulator.
	  Development of FEL power along the undulator was calculated
	  in a layout including a 6~m undulator and a matched beam for it, assuming realistic beam quality.
}
	\label{fig:longund}
\end{center}
\end{figure}

However, 
the slippage in one saturation length
becomes 5.8~ps, which is much larger than the bunch length.
In view of this, the above steady-state estimation needs a major correction.
We moved to a time-dependent simulation that includes the slippage effect with a finite bunch length
assuming a Gaussian-shaped bunch.
Figure \ref{fig:longund} plots the FEL power development along the undulator.
It shows that more than 5~m of undulator length
 is necessary to reach the saturation point.
Considering practical issues, such as cost and lead time, in fabricating one long undulator, 
we decided to use two 3-m undulators.

%%%%%%%%%%%%%%%%%%%%%%
%%%%%%%%%%%%%%%%%%%%%%

%---------------------------------------------
\subsection{Simulation of SASE FEL}
%---------------------------------------------

A detailed description of the layout will be given in Sec. \ref{sec:setup} and \cite{cerl_construction}.
For the start-to-end simulation, i.e., from the cathode of the gun to the end of FEL undulator including the development of FEL radiation,
we used three simulation codes.
The injector part where the beam energy is low
was implemented using GPT \cite{gpt}.
After the main linac, 
the particle distribution was passed to elegant \cite{ele},
which was suitable for tracking the beam transport system
up to the entrance of the FEL undulator.
The distribution was then used as the input of FEL simulation by GENESIS \cite{genesis}.
The final calculation was performed with 500k macro particles.

\begin{figure}[htb]
	\begin{center}
 	 \includegraphics[bb=0 0 550 450, width= 0.8\linewidth]{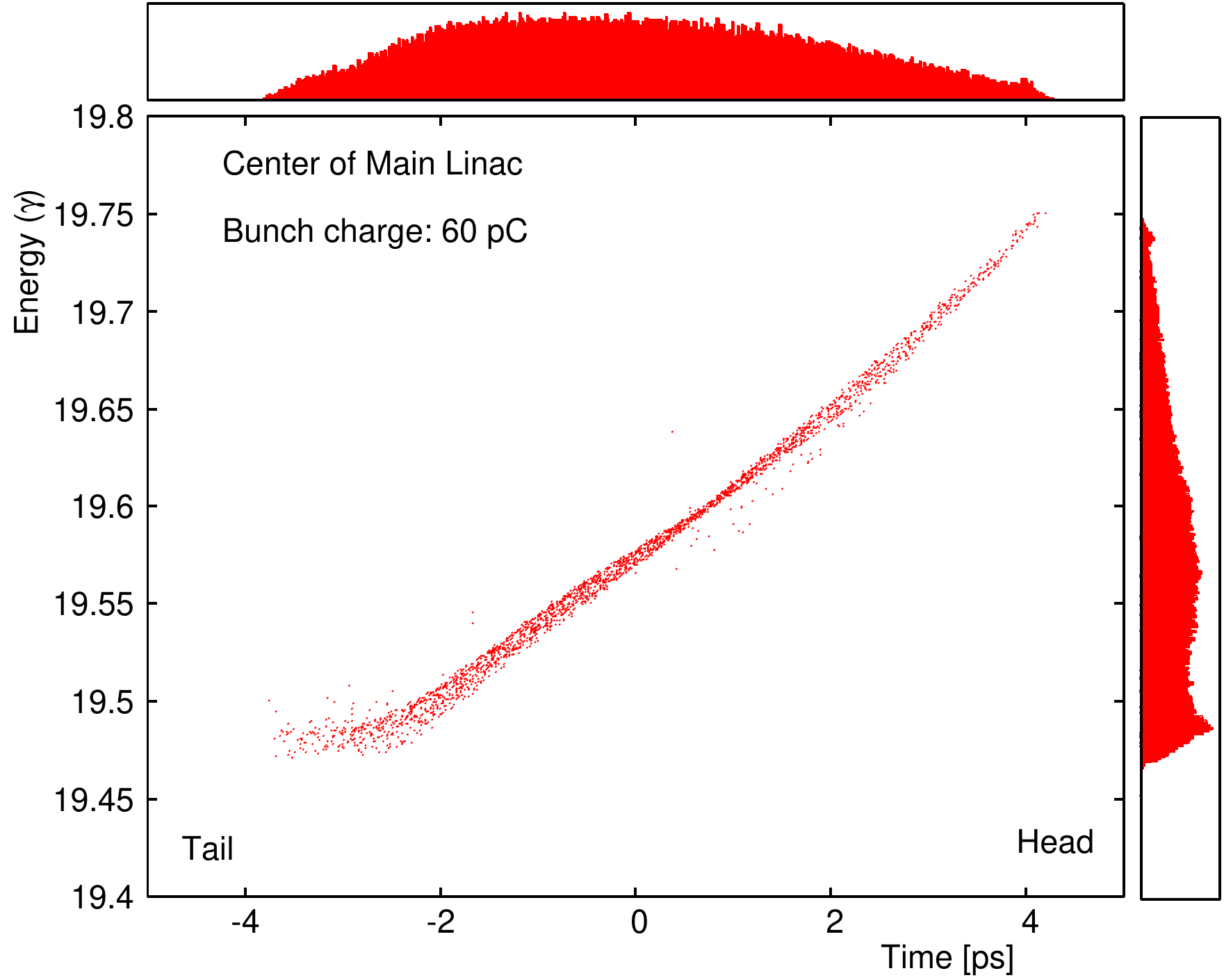}
	  \caption{
	  Simulated particle distribution at the center of the main linac using GPT.
	  The particle distribution in the longitudinal phase space of time and energy is shown.
	  The shapes projected on each axis are also shown.
}
	\label{fig:injectorsim_pointD}
\end{center}
\end{figure}

The particle distribution of the optimized injector setting
was simulated using GPT.
The simulated particle distribution at the entrance of the second main linac cavity
was used as the starting point in the elegant simulation.
Figure \ref{fig:injectorsim_pointD} shows 
the longitudinal phase space distribution.
It is clear that the beam has an energy chirp in the bunch.

%%%%%%%%%%%%%%%%%%%%%%%
%%%%%%%%%%%%%%%%%%%%%%%

From the entrance of the second main linac,
the tracking simulation was performed using elegant.
It includes the longitudinal space charge effect (LSC) and the coherent synchrotron radiation effect (CSR),
but the transverse space charge effect (TSC) was ignored.
The TSC is not negligible; in this case, however, 
it can be locally corrected as a perturbation from the viewpoint of Twiss parameter tuning
as explained in \ref{sec:sc}.
In this tracking,
the procedure of beam handling for bunch compression, i.e., 
the RF phase of the main linac and
the longitudinal dispersion ($R_{56}$ ) tuning in the arc section, was included.
$R_{56}$ is defined by $R_{56}=d l / d\delta$, 
i.e.,
the path length ($l$) difference depending on the relative energy ($\delta$) variation.

The FEL simulation was performed using the time-dependent calculation of GENESIS.
The layout consists of two undulators and the space between them.
The simulations were conducted in three steps.
The first step was for the first undulator
using the particle distribution received from elegant as the initial distribution.
To reproduce the start-up effect due to the fine density distribution
that may not appear in the particle distribution,
we employed an empirical parameter to set the initial energy modulation.
In the second step,
using the particle distribution at the exit of the first undulator,
the tracking calculation in the space between the undulators was performed.
At this time, the radiation data were removed
because they do not overlap with the electron bunch in time
after the space.
In the third step,
the simulation of the second undulator was conducted
using the particle distribution above as the input.
The radiation began growing from zero.

Although we performed many simulations in the designing stage,
the actual parameters realized in the beam experiments have changed
after the beam tuning explained in Sec. \ref{sec:experiment}.
Here, we show the simulation results for the beam parameters corresponding to the experimental case.
The parameters used in the simulation
are summarized 
in Table \ref{tab:simparam}.

%%%%%%%%%%%%%%%%%%%%%
\begin{table}[htb]
  \begin{center}
    \caption{Parameters used in the simulation}
    \label{tab:simparam}
    \begin{tabular}{l c | l } \hline
      Initial distribution & & \\ \hline
      Bunch charge & & 60~pC\\ 
      Norm. Emittance (X)& & 1.63~$\mu$m \\
      Norm. Emittance (Y)& & 2.01~$\mu$m \\
      Bunch length & & 1.85~ps  (RMS)\\
      Energy spread &  $\Delta \gamma / \gamma$ & $2.23\times10^{-3}$ (RMS) \\ \hline
      Transport & & \\ \hline     
     Chirp phase & & +29$^\circ$ (at 1.3~GHz)\\
      $R_{56}$ & & -0.282~m \\ \hline
      Entrance of Undulator & & \\ \hline
      Energy & $\gamma$ & 34.46\\
      Norm. Emittance (X)& & 4.45~$\mu$m \\
      Norm. Emittance (Y)& & 1.85~$\mu$m \\     
      Bunch length & & 1.94 ~ps  (RMS)\\
      Energy spread & $\Delta \gamma / \gamma$& $6.22 \times 10^{-3}$\\
      Energy modulation & & 0.02\% \\ \hline
      Undulator & & \\ \hline
      Strength & $a_w$  & 0.97 \\
      Tapering coefficient & & -0.048 \\
      Period &$\lambda_u$ & 24 mm \\ 
      Length & & 3 m $\times $ 2\\
      
             \hline
  \end{tabular}
  \end{center}
\end{table}
%%%%%%%%%%%%%%%%%%%%%%

\begin{figure}[htb]
	\begin{center}
 	 \includegraphics[bb=0 0 600 400, width= 0.8\linewidth]{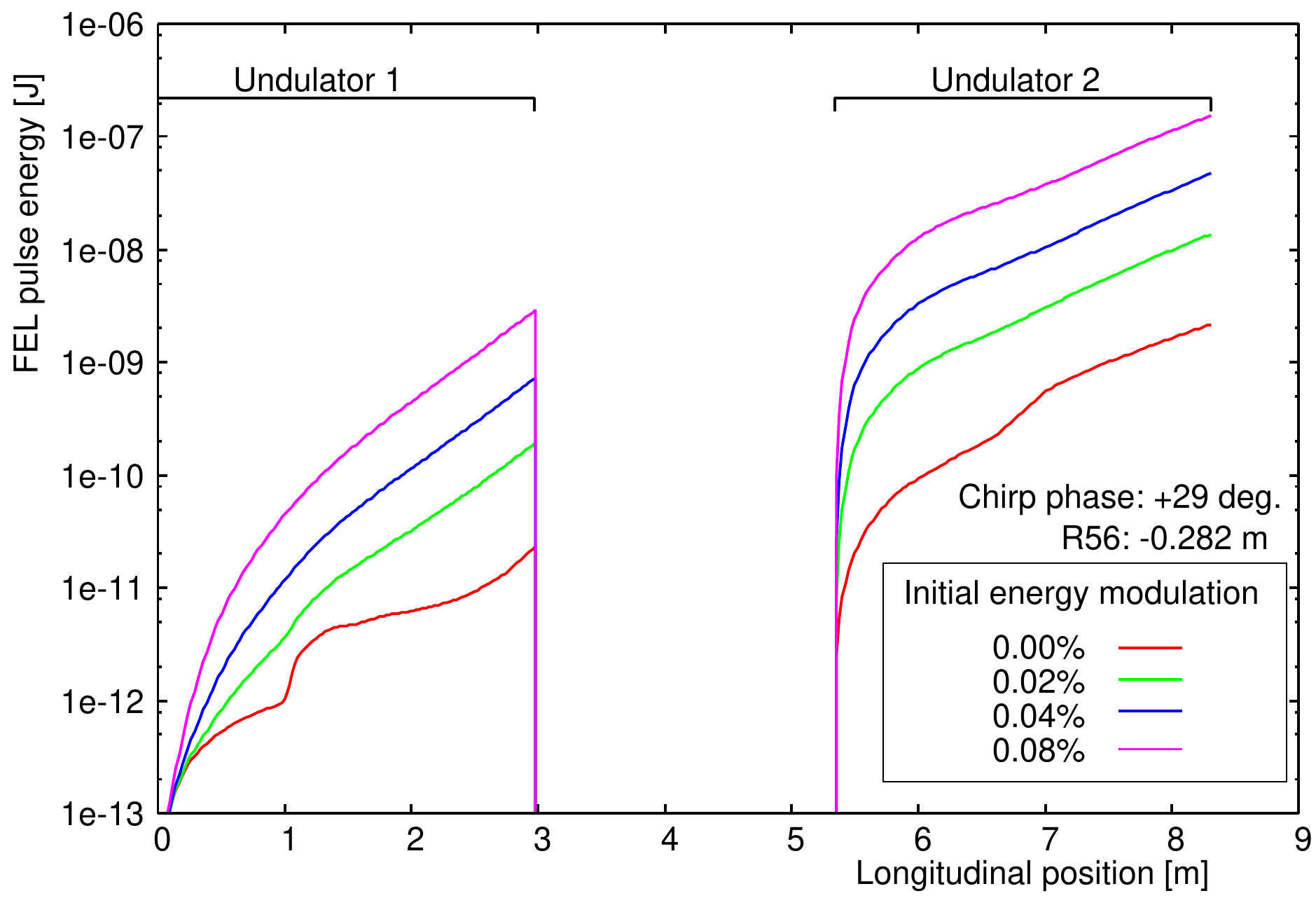}
	  \caption{
	  Simulated FEL power development along the undulator
	  based on the beam parameters determined in the experiment,
	  chirp phase of +29$^\circ$ and $R_{56}$ of $-0.282$~m.
	  The initial energy modulation is an empirical parameter 
	  to adjust the absolute power using the start-up effect.
	  The radiation wavelength was set to 20 $\mu$m.
}
	\label{fig:powersim}
\end{center}
\end{figure}

Figure \ref{fig:powersim} shows the simulation results of
FEL output power along the longitudinal position.
The start-up can be changed by the energy modulation we set empirically.
The gain at the second undulator is estimated as approximately 10 for any case.

%%%%%%%%%%%%%%%%%%%%%%%
%%%%%%%%%%%%%%%%%%%%%%%

\begin{figure}[htb]
	\begin{center}
 	 \includegraphics[bb=0 0 550 450, width= 0.8\linewidth]{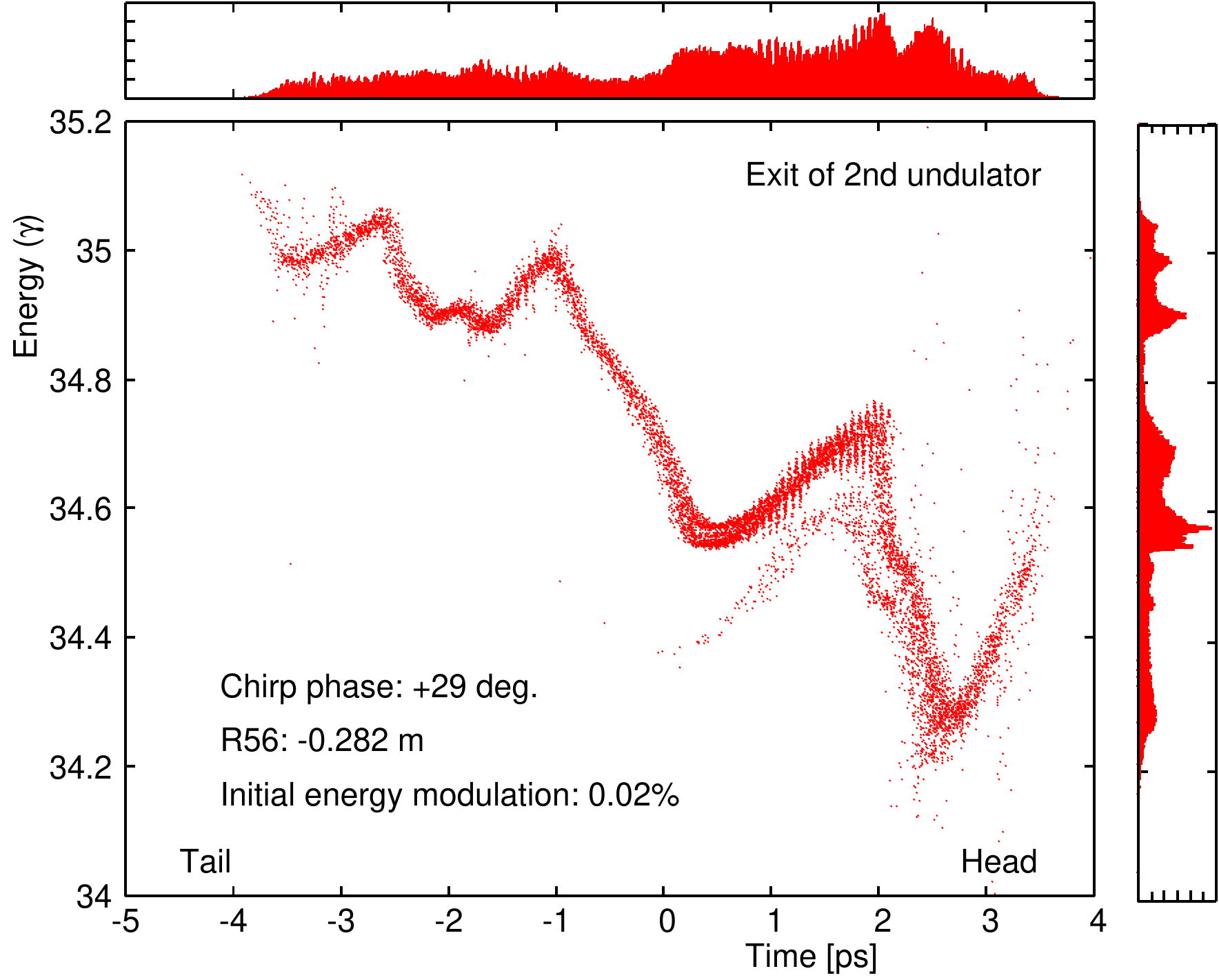}
	  \caption{
	  Longitudinal phase space distribution at the exit of the undulator.
	  The shapes projected on each axis are also shown.
	  Fine structures corresponding to the radiation wavelength
	  are formed by FEL interaction,
	  showing that some localized parts contribute to the FEL gain.
	The radiation wavelength was set to be 20 $\mu$m in this calculation.
}
	\label{fig:simslice}
\end{center}
\end{figure}

The particle distribution in the longitudinal phase space at the exit of the second undulator
is shown in Fig. \ref{fig:simslice}.
Due to the microbunching instability,
the phase space distribution has a complicated structure with kinks.
When strong FEL interaction has occurred,
the energy of the electron beam should be modulated at the radiation wavelength.
Such a fine structure of energy modulation appears at several locations in the bunch.
It appeares that some localized parts of the bunch independently contribute to the FEL gain.
If the whole bunch could contribute to the growth of a single spike of radiation pulse,
the efficiency would become much higher.
However, in the given scenario of a complicated distribution with a large energy spread,
the signal can grow only locally.
This simulation case shows that
even if the situation is not ideal,
there is still a chance to obtain an FEL gain.

%%%%%%%%%%%%%%%%%%%%%%%
%%%%%%%%%%%%%%%%%%%%%%%

\begin{figure}[htb]
	\begin{center}
 	 \includegraphics[bb=0 0 600 400, width= 0.8\linewidth]{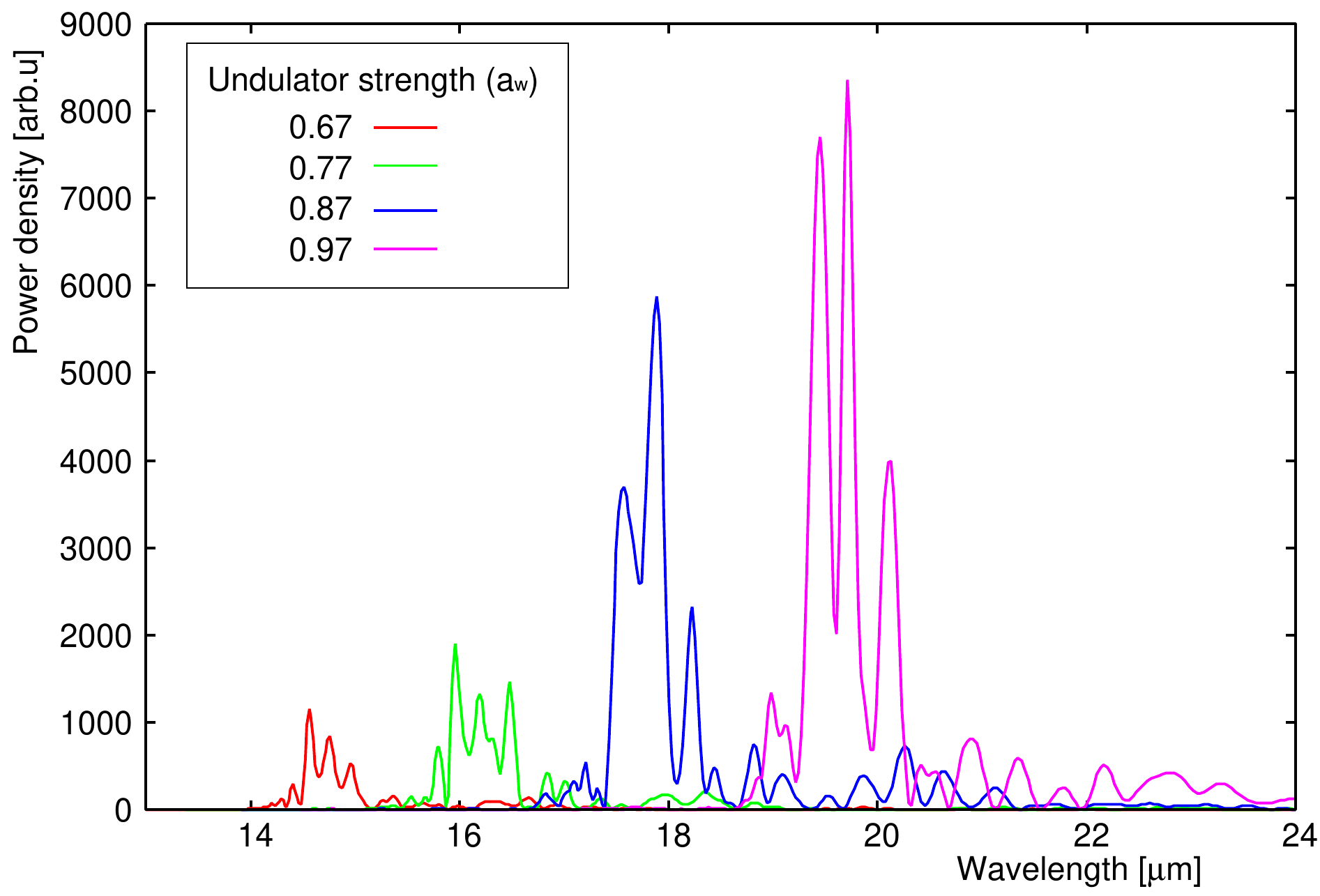}
	  \caption{
	  Simulation results of the radiation spectrum.
	  Four cases of undulator strength are shown.
}
	\label{fig:simspectrum}
\end{center}
\end{figure}

\begin{figure}[htb]
	\begin{center}
 	 \includegraphics[bb=0 0 600 400, width= 0.8\linewidth]{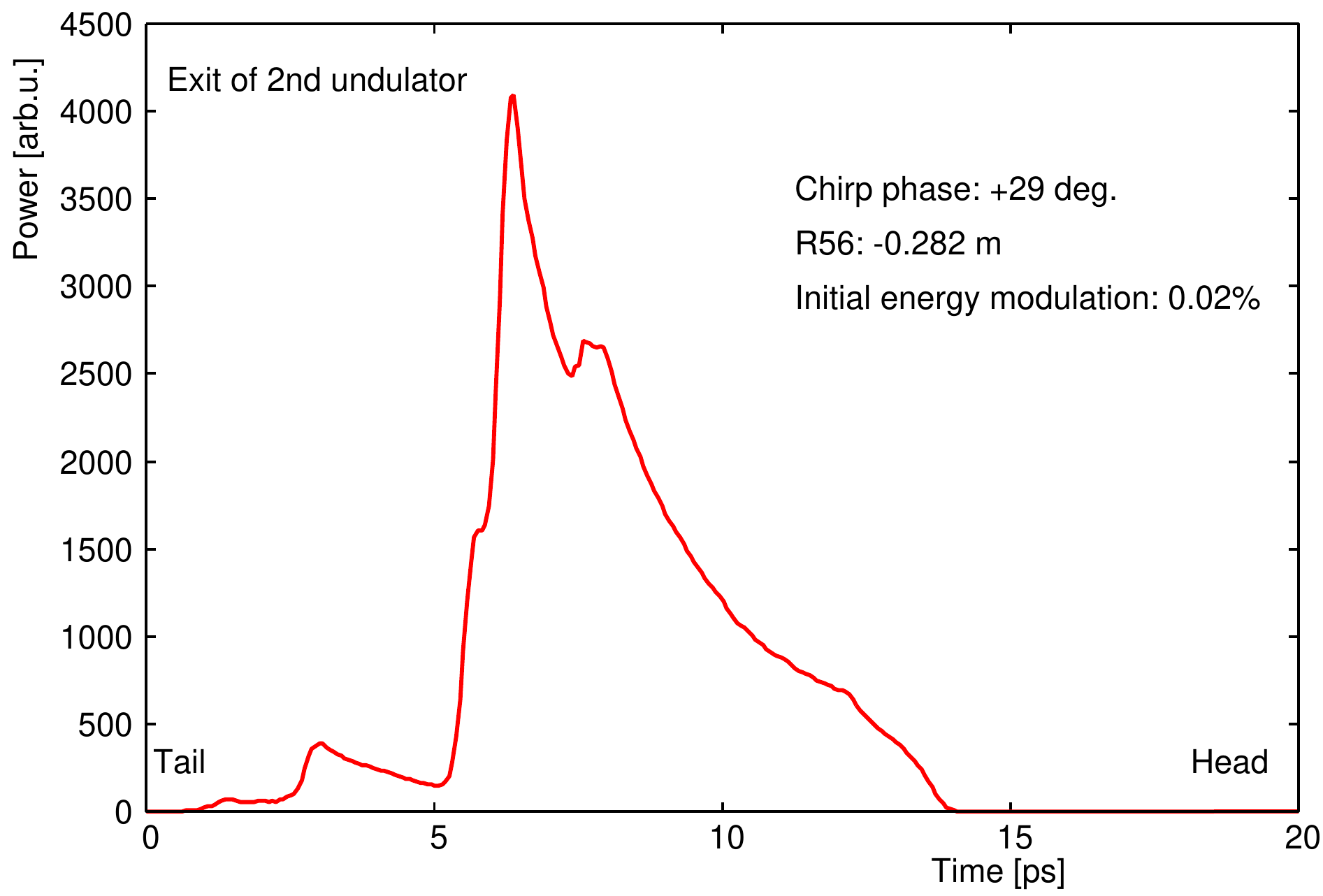}
	  \caption{
	  Simulated radiation pulse shape at the exit of the second undulator.
	  The pulse shape contains several (more than three) peaks.
	  Each of the peaks appears to originate from a different part of the bunch.
	  The radiation wavelength was set to 20 $\mu$m in this calculation.
}
	\label{fig:pulseshape}
\end{center}
\end{figure}

The radiation spectrum can be obtained by Fourier analysis of the radiation field data extracted from the simulation.
Figure \ref{fig:simspectrum} shows the results.
The bandwidth was estimated as 0.8~$\mu$m in full width at half maximum (FWHM).
The center wavelength can be changed by the undulator strength.
The spectrums show multiple peaks,
which may be attributed to the local power growth 
explained in Fig. \ref{fig:simslice}.
%%%%%%%%%%%%%%%%%%%%%%%
%%%%%%%%%%%%%%%%%%%%%%%
%
Figure \ref{fig:pulseshape} shows the simulated radiation pulse shape.
It turns out  not to be a simple single spike.
It appeares that the signal is composed of several pulses that have independently grown at different parts in a bunch.

%%%%%%%%%%%%%%%%%%%%%%%
%%%%%%%%%%%%%%%%%%%%%%%

\begin{figure}[htb]
	\begin{center}
 	 \includegraphics[bb=0 0 350 300, width= 0.6\linewidth]{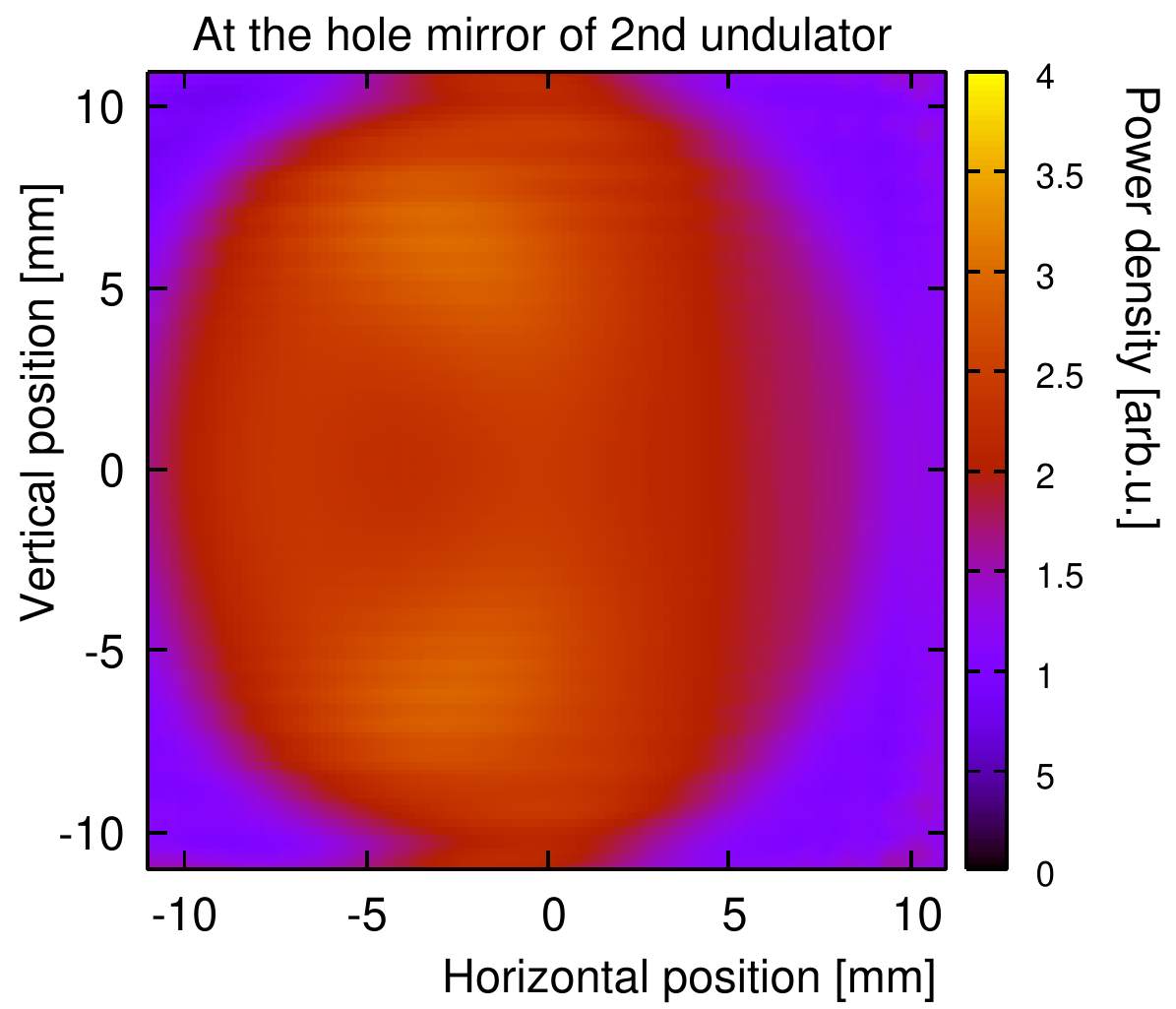}
	  \caption{
	  Spatial profile of radiation power density 
	  at the extraction hole mirror of the second undulator.
	  The radiation wavelength was set to be 20 $\mu$m in this calculation.
}
	\label{fig:simprofile}
\end{center}
\end{figure}

Figure \ref{fig:simprofile} shows the spatial power distribution 
at 1.08~m downstream from the exit of the second undulator,
which corresponds to the position of the extraction mirror.
The diameter of the profile is approximately 20~mm.

%%%%%%%%%%%%%%%%%%%%%%%
%%%%%%%%%%%%%%%%%%%%%%%

%===========================================
\section{EXPERIMENTAL SETUP}
\label{sec:setup}
%===========================================

%---------------------------------------------------
\subsection{Beam transport system}
%---------------------------------------------------

\begin{figure}[htb]
	\begin{center}
 	 \includegraphics[bb=0 0 1500 800, width= 0.8\linewidth]{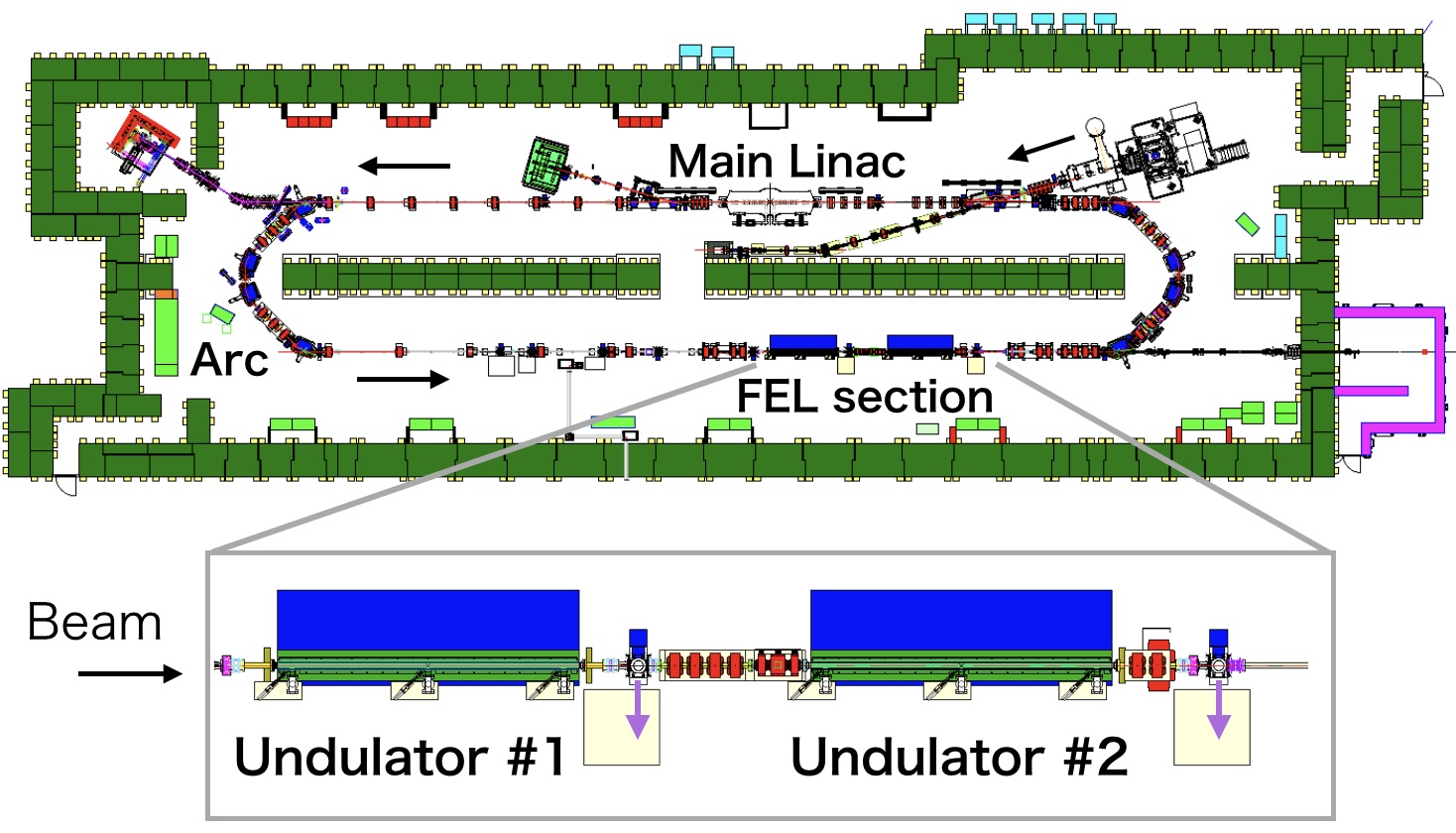}
	  \caption{
	  Layout of cERL.
	  The electron beam from the injector 
	  is first accelerated by the main linac.
	  Next, it is transported to the return loop through the arc section
	  to the newly constructed FEL section.
}
	\label{fig:cerllayout}
\end{center}
\end{figure}

Since the initial construction  \cite{cerl_construction},
the cERL has been operating as a test accelerator 
with its layout being changed in every few years \cite{cerl_akagi, cerl_rcdr}.
Figure \ref{fig:cerllayout} shows the layout for the FEL experiment.
The downstream half of the straight line in the return loop
was modified.

%%%%%%%%%%%%%%%%%%%%%%%
%%%%%%%%%%%%%%%%%%%%%%%
%

%%%%%%%%%%%%%%%%%%%
\begin{figure}[htb]
	\begin{center}
 	 \includegraphics[bb=0 0 500 200, width= 0.9\linewidth]{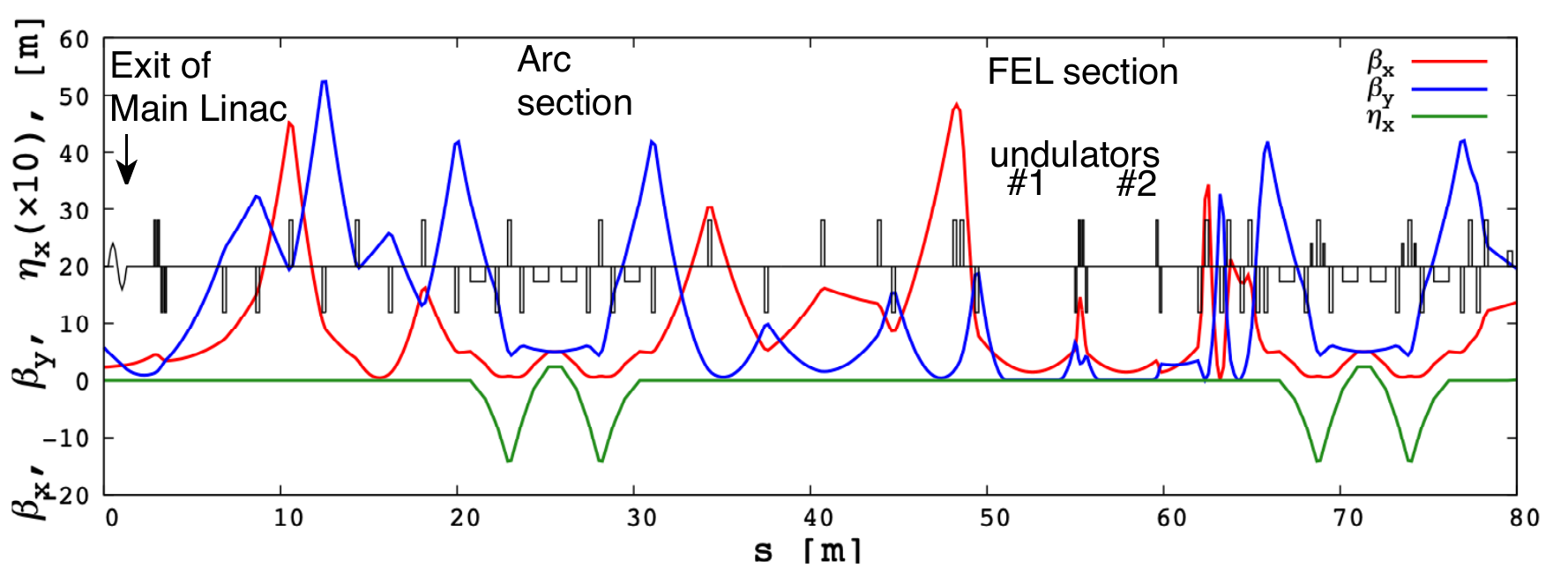}
	  \caption{
	  Designed beam optics of the return loop,
	  starting from the exit of the main linac to the exit of the second arc.
	  The beta functions of the horizontal and vertical plane and the dispersion function of horizontal plane
	  are shown.
}
	\label{fig:circoptics}
\end{center}
\end{figure}
%%%%%%%%%%%%%%%%%%%%

The designed beam optics of the return loop is shown in Fig. \ref{fig:circoptics}.
This beam optics was calculated starting 
from the Twiss parameters at the exit of the main linac,
and transported by linear optics without any collective effects.
The arc section consists of four 45$^\circ$ bending magnets,
six quadrupoles, and two sextupoles.
The linear optics was designed to be mirror symmetric about the center of the arc.
It was designed to be achromat and to be $R_{56}$ controllable
by changing the strength of the quadrupole magnets in the arc section.

For the FEL beam tuning, where the longitudinal density distribution in the bunch is important,
the $R_{56}$ tuning is a sensitive knob that optimizes the FEL output.
The details of the design and tuning of bunch compression 
will be given in elsewhere \cite{cerl_shimada}.
For fine beam optics tuning to the undulators, 
four quadrupole magnets upstream of each undulator were chosen
to control four degrees of freedom.
They were used as combined knobs, as explained in \ref{sec:knobs}.

%---------------------------------------------------------
\subsection{Layout of FEL  section}
%---------------------------------------------------------

\begin{figure}[htb]
	\begin{center}
 	 \includegraphics[bb=0 0 720 250, width= 0.9\linewidth]{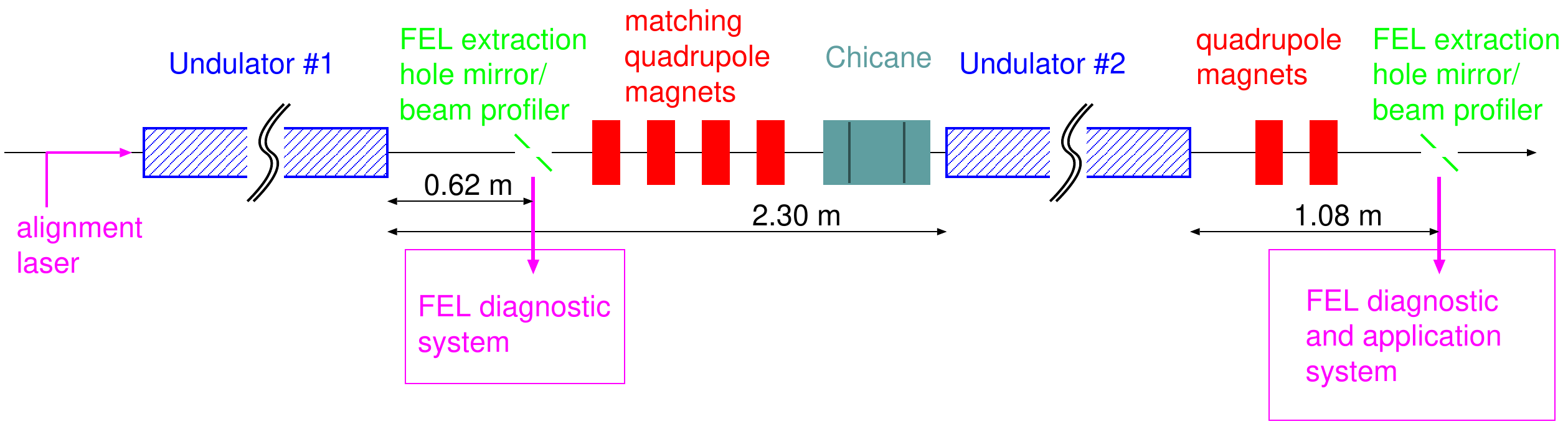}
	  \caption{
	  Layout of the FEL section.
	  A FEL extraction mirror, matching quadrupoles, and a chicane magnet are installed
	  in the 2.30~m space between the two 3-m undulators.
	  The second FEL extraction mirror is installed 
	  behind of two quadrupole magnets.
}
	\label{fig:undsection}
\end{center}
\end{figure}

The layout of the FEL section is shown in Fig. \ref{fig:undsection}.
Two 3-m undulators were located with a separation of 2.3~m.
Various components were installed in the short space between the undulators,
The installed components were a hole mirror system to extract radiation 
and that can also be used as a beam profile monitor,
four quadrupole magnets for optics matching to the second undulator,
and a chicane magnet that could help enhance the FEL power with the optical klystron effect \cite{ok}.

%%%%%%%%%%%%%%%%%%%
\indent
%%%%%%%%%%%%%%%%%%%

We adopt a planar-type undulator in the horizontal plane,
utilizing the adjustable phase scheme (APU \cite{apu}).
The effective strength of the vertical magnetic field can be varied
by sliding the relative position 
(we call it as phase by defining the position shift of one period to be $2\pi$) of the girder of each side
while the mechanical gap is fixed as
\begin{equation}
a_w = a_{w0} \cos(\textrm{phase}/2) \quad.
\end{equation}
In the case of low beam energy,
the vertical focusing effect of the undulator becomes strong.
It is necessary to carefully match the Twiss parameter to maintain the small beam size throughout the undulator,
and also to connect the beam condition downstream.
The vertical motion of the beam is represented as 
an oscillation equation:
\begin{equation}
\frac{d^2 y }{ds^2} = - \left( \frac{2 \pi a_{w0} }{\lambda_u \gamma}\right)^2 y \quad ,
\end{equation}
where the $a_{w0}$ is the undulator strength at the maximum effective field.
The focusing effect does not change
even if the effective field is reduced by sliding the poles of the APU.
This feature is an advantage of the APU, especially for the ERL layout,
where the beam condition in downstream must be controlled well
while changing the FEL wavelength.

\begin{figure}[htb]
	\begin{center}
 	 \includegraphics[bb=0 0 4600 3300, width= 0.7\linewidth]{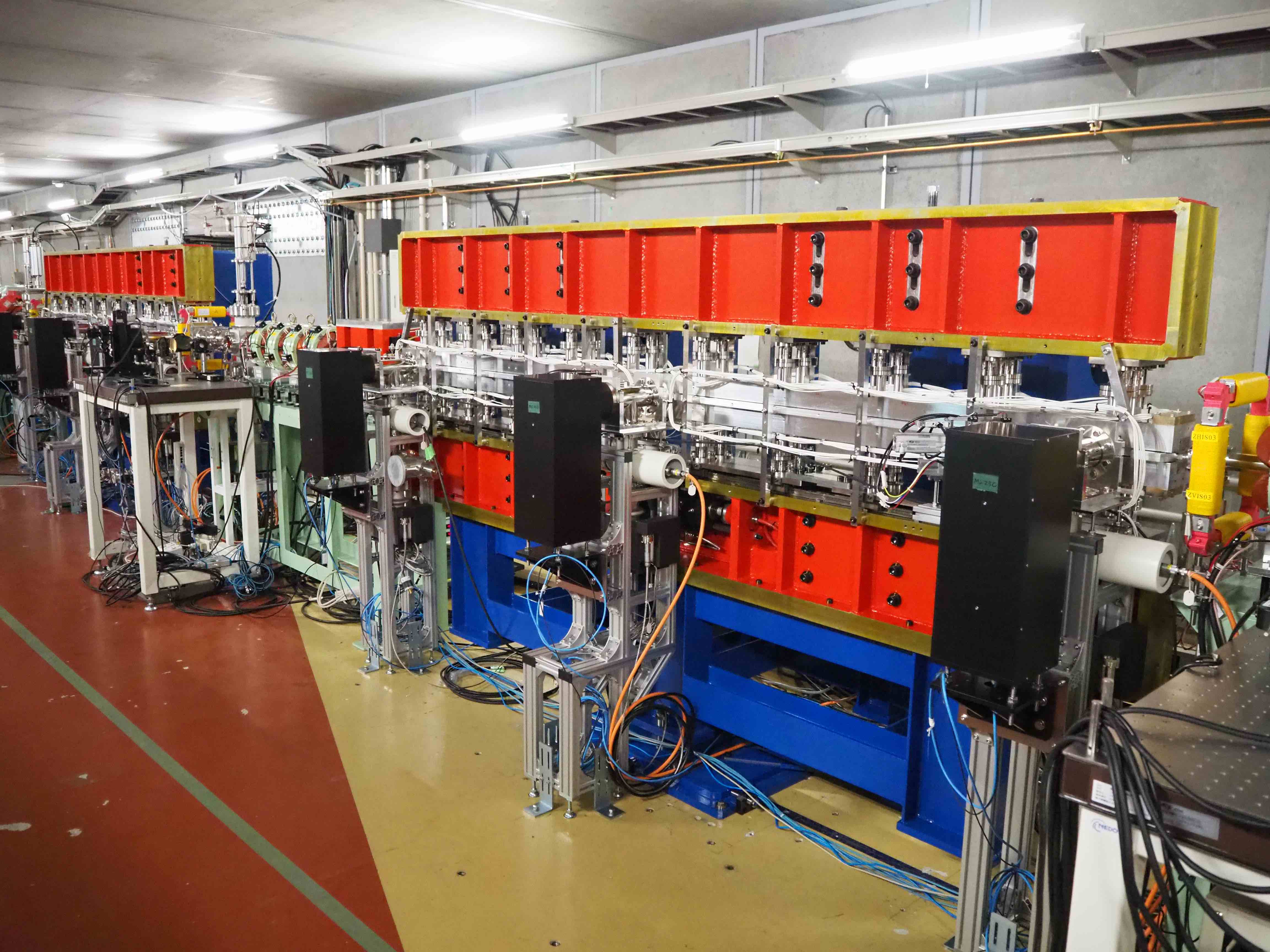}
	  \caption{
	  Undulators viewed from downstream of the second undulator.
	  The three boxes placed at the side of each undulator are for the beam profile monitor systems.
}
	\label{fig:und_pic}
\end{center}
\end{figure}

\indent
The undulators were newly fabricated for this experiment.
Figure \ref{fig:und_pic} shows the experimental setup.
To obtain the magnetic field strength for the undulator period 
that corresponds to the target radiation wavelength at the given beam energy,
a relatively small gap of 10.0~mm was required.
The undulator was modified so that the magnetic field strength was tapered along the beam axis.
The tapering was achieved by inclining the whole upperside girder on the support structure
to obtain a linear variation in the gap.
The tapering coefficient was designed to be $-0.048$,
which means that the field strength at the exit was higher by the ratio with respect to that at the entrance.
Both the undulators were modified to have the same tapering coefficient.
The parameters of the undulator are summarized in Table \ref{tab:undparam}.
%The details of the undulator are reported in \cite{und_adachi}.\\
%%%%%%%%%%%%%%%%%%%%%
\begin{table}[htb]
  \begin{center}
    \caption{Parameters of the undulator}
    \label{tab:undparam}
    \begin{tabular}{l c | l } \hline
      Type & & Planar (APU)\\ 
      Gap & & 10.0~mm (fixed)\\
      Strength (maximum) & $a_w$ & 0.97  ($=K/\sqrt{2}$) \\ 
      Tapering coefficient & & -0.048  \\ 
      Material  &  & Nd-Fe-B \\
      Number of period &  & 124 \\
      Period & $\lambda_u$ & 24~mm \\
      Number of unit & & 2\\
      \hline
  \end{tabular}
  \end{center}
\end{table}
%%%%%%%%%%%%%%%%%%%%%%
%%%%%%%%%%%%%%%%%%%
\indent
%%%%%%%%%%%%%%%%%%%

\begin{figure}[htb]
	\begin{center}
 	 \includegraphics[bb=0 0 650 200, width= 0.8\linewidth]{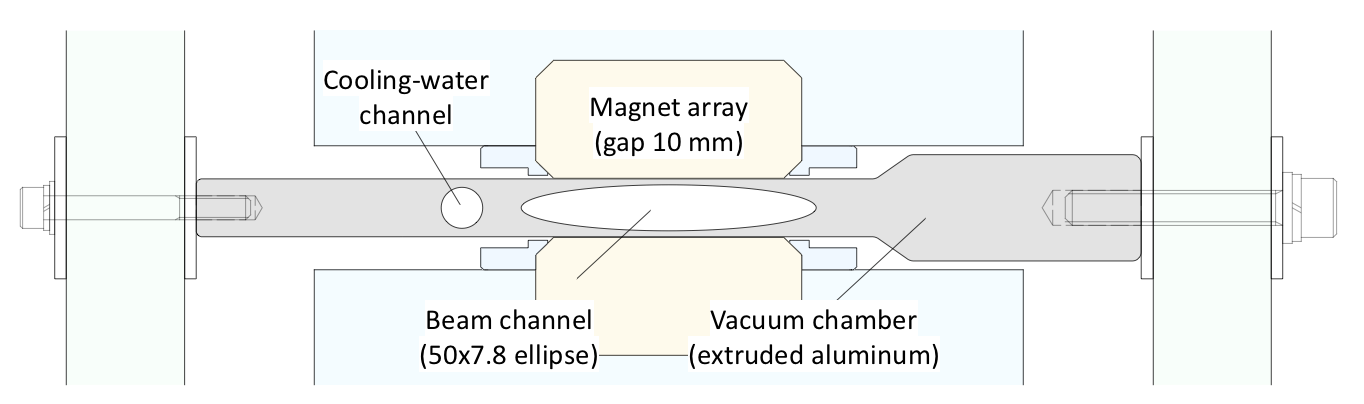}
	  \caption{
	  Cross section of the undulator and the vacuum duct.
}
	\label{fig:chamber_cross}
\end{center}
\end{figure}

\begin{figure}[htb]
	\begin{center}
 	 \includegraphics[bb=0 0 520 350, width= 0.8\linewidth]{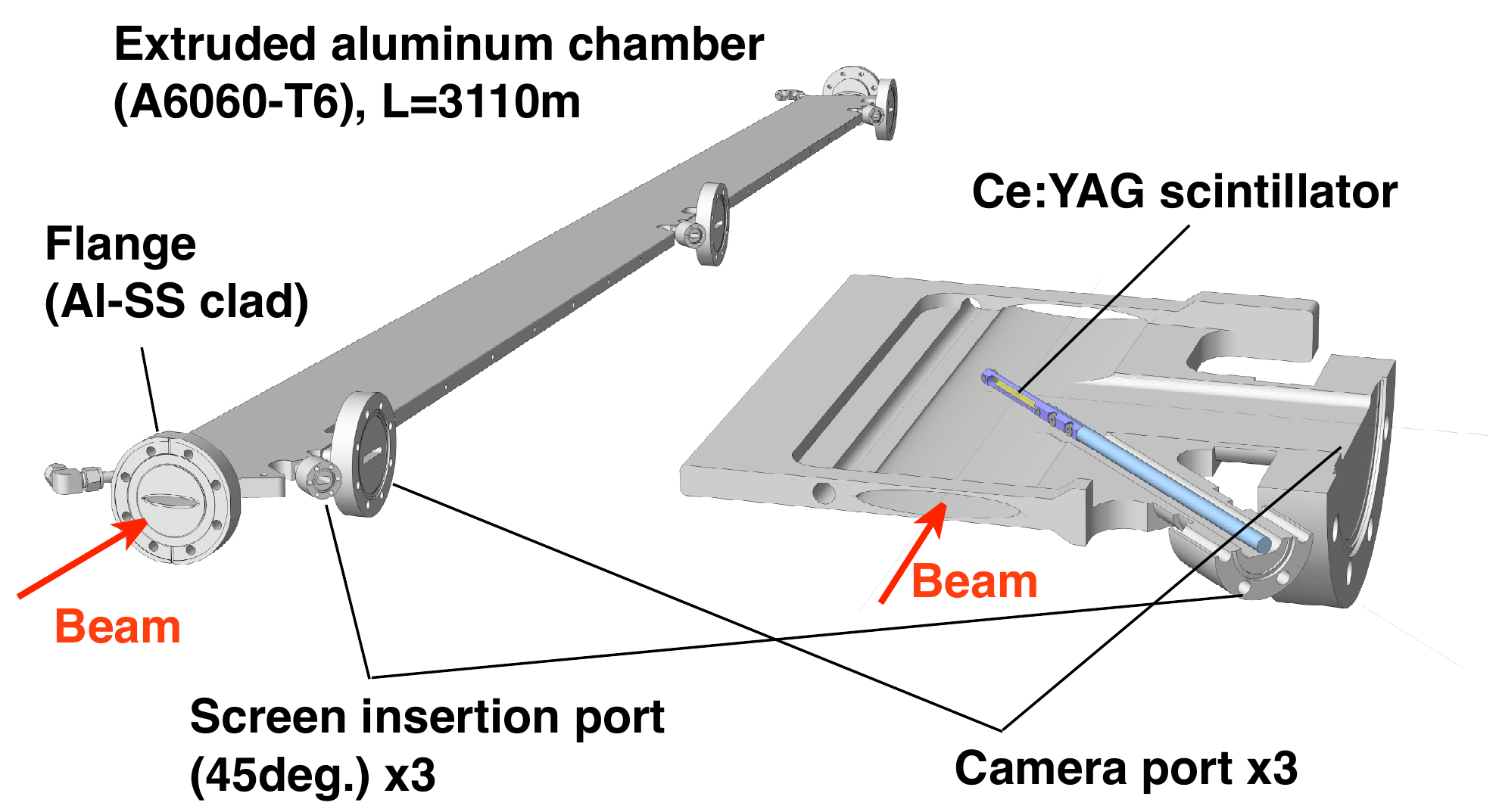}
	  \caption{
	  Structure of the vacuum duct in the undulator.
	  It includes ports for three beam profile monitors.
	  Each beam profile monitor part consists of a 45$^\circ$ screen insertion port
	  and a camera viewing port.
	  (Based on production drawings of SAES Getters S.p.A. and SEAS RIAL Vacuum S.r.l. )
}
	\label{fig:chamber}
\end{center}
\end{figure}

The small gap of the undulators
required the vacuum duct in the undulators 
to have a flat cross section with thin walls,
as shown in Fig. \ref{fig:chamber_cross}.
The duct is made of aluminum alloy fabricated by the extruded technique. 
The inner cross section of the duct
has an elliptical shape with diameters of 50~mm and 7.8~mm  in the horizontal and vertical directions, respectively.
%%%%%%%%%%%%%%%%%%%
\indent
%%%%%%%%%%%%%%%%%%%
%
Figure \ref{fig:chamber} shows the structure of the undulator duct.
Beam profile monitors,
each of which consists of a scintillator insertion system and a window to monitor it using a camera system,
were installed at three locations in each of the undulator ducts.
The profile monitors were located
at the center of the undulator 
and 0.15~m from both ends of the undulator pole edge.
The scintillators are Ce-doped YAG crystal of 0.1~mm thickness.
The effective area of the scintillators is
30~mm $\times$ 4~mm and they have a racetrack shape.
They are inserted in 45$^\circ$ with respect to the beam direction,
and are viewed from the rear.

%%%%%%%%%%%%%%%%%%%
\indent
%%%%%%%%%%%%%%%%%%%

\begin{figure}[htb]
	\begin{center}
 	 \includegraphics[bb=0 0 750 430, width= 0.8\linewidth]{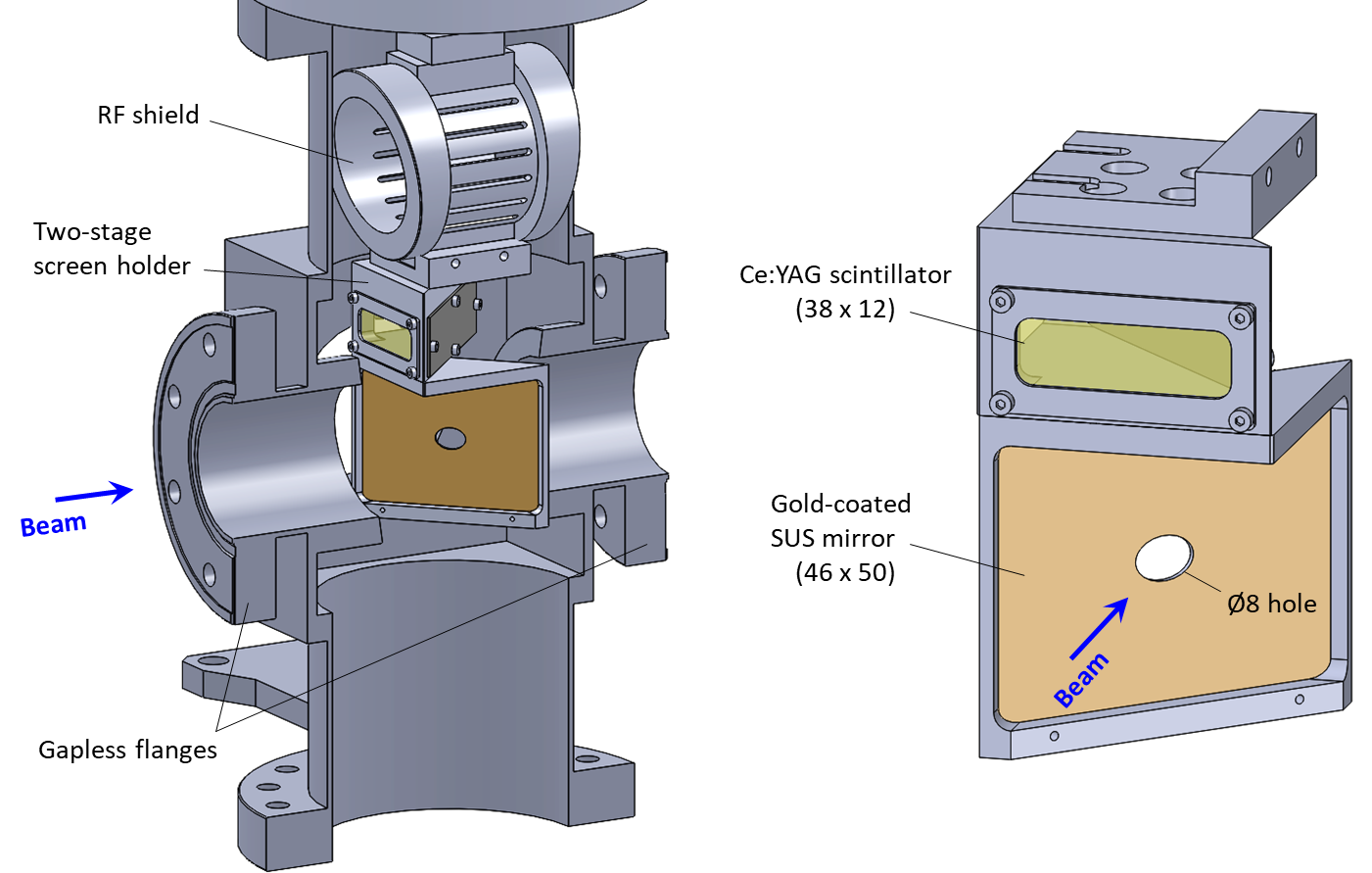}
	  \caption{
	  Structure of the insertion system for the FEL extraction hole mirror system.
	  The gold-coated mirror with a hole of 8~mm diameter is inserted on the beam axis
	  at 45$^\circ$ to reflect the FEL radiation in the perpendicular direction.
	  The system can be inserted in two positions.
	  The other position is for the Ce:YAG scintillator to measure the electron beam.
}
	\label{fig:holemirror}
\end{center}
\end{figure}

An FEL radiation output system was installed after each undulator.
Radiation was extracted to the air through the KRS-5 window
by reflecting it in the perpendicular direction with a hole mirror as shown in Fig. \ref{fig:holemirror}.
A gold-coated SUS plate of dimensions 46~mm $\times$ 50~mm with a circular hole of 8~mm diameter viewed from  the beam
was inserted on the beam axis.
According to the simulation of Fig. \ref{fig:simprofile},
95\% of the radiation power can be extracted
without destroying the electron beam.
The hole mirror can be switched to a beam profile monitor 
which is a Ce:YAG scintillator viewed from the rear through an SUS mirror
so that the electron beam size can be measured.
A camera system monitors the setup from the window at the opposite side of the radiation extraction.

%%%%%%%%%%%%%%%%%%%
\indent
%%%%%%%%%%%%%%%%%%%

To match the beam optics to the second undulator,
four quadrupole magnets of 100~mm thickness were located between the undulators.
They were set in a Defocus-Focus-Focus-Defocus configuration
for the horizontal plane,
it actually composed a triplet optics.

\begin{figure}[htb]
	\begin{center}
 	 \includegraphics[bb=0 0 500 250, width= 0.7\linewidth]{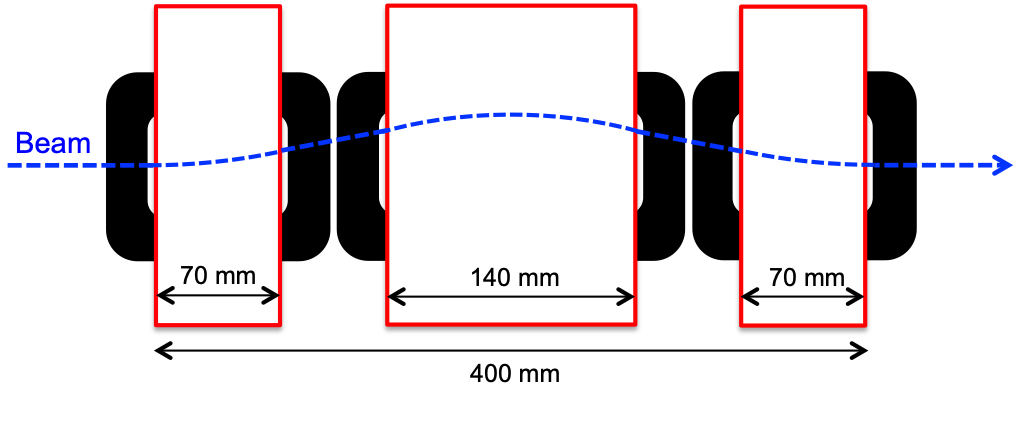}
	  \caption{
	  Chicane magnet.
	  Three dipole magnets produce a closed bump in the horizontal plane.
}
	\label{fig:chicane}
\end{center}
\end{figure}

A chicane bump system consisting of three dipole magnets
(as explained in Fig. \ref{fig:chicane})
was also installed between the undulators.
The role of the chicane in a typical FEL is as a phase shifter
to match the phase of the radiation between the undulators.
However, its role is different in our case,
where radiation from the first undulator totally slips the bunch duration, and there is no overlap
at the second undulator.
The chicane was installed in anticipation of 
the effect of the optical klystron; i.e.,
the density modulation initiated by the energy modulation after the first undulator
could be enhanced,
which could help increase the FEL output.
The maximum bump height within the fiducial volume is 23~mm,
which corresponds to an  $R_{56}$ of 6~mm.

%%%%%%%%%%%%%%%%%%%
\indent
%%%%%%%%%%%%%%%%%%%

%---------------------------------------------------------
\subsection{FEL diagnostic system}
%---------------------------------------------------------

\begin{figure}[htb]
	\begin{center}
 	 \includegraphics[bb=0 0 500 600, width= 0.8\linewidth]{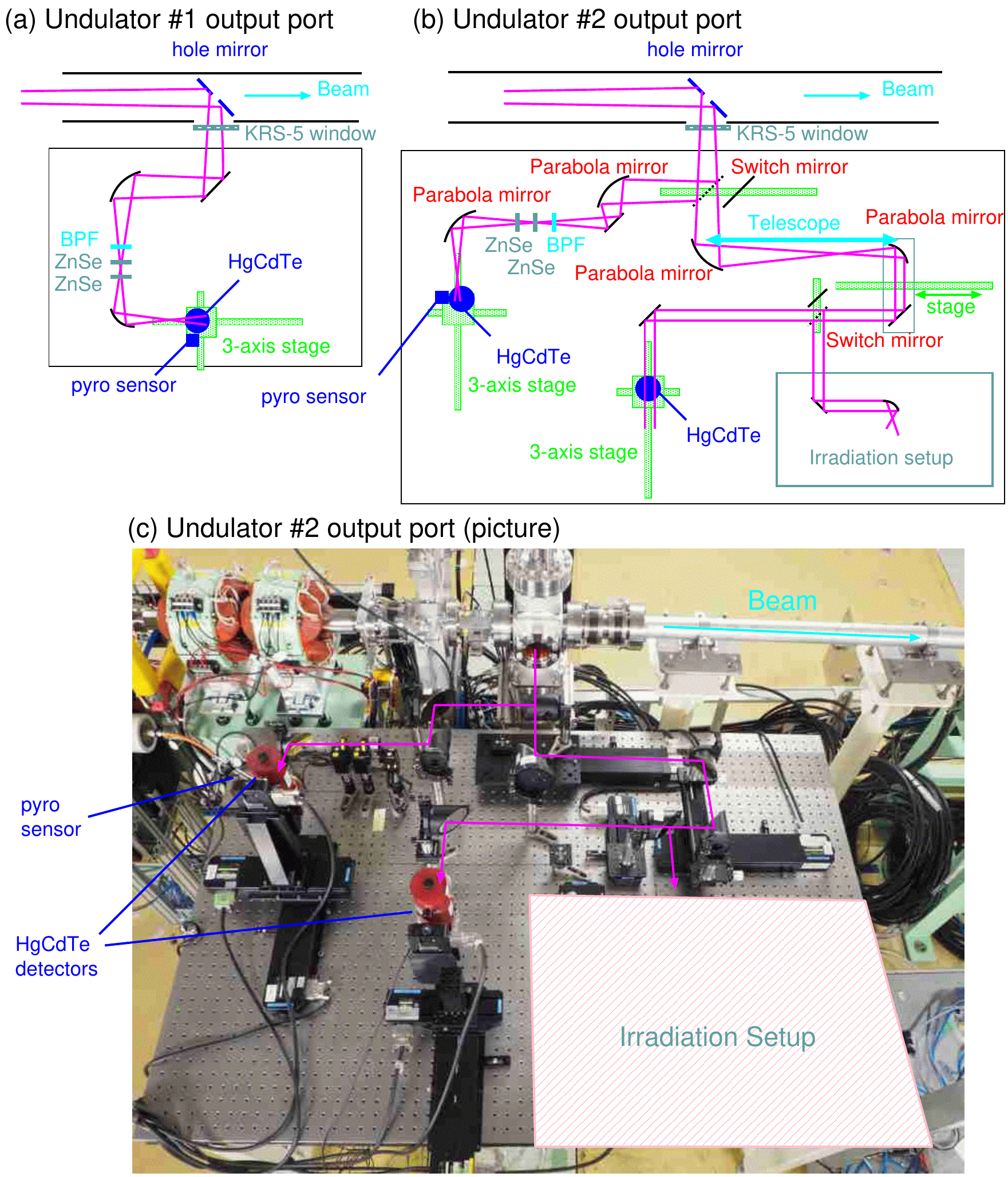}
	  \caption{
	  Setup of the radiation diagnostic system.
	  The basic design of the monitor system is the same.
	   It consists of focusing optics with a parabolic mirror,
	   three filters placed at the focal point,
	   and another parabolic mirror to focus on the detectors on a three-axis stage.
	   The system at the second undulator has the irradiation system behind the switch mirror.
}
	\label{fig:udetector_setup}
\end{center}
\end{figure}

The radiation output port downstream of the first undulator was prepared primarily for beam tuning.
Figure \ref{fig:udetector_setup}(a) shows the layout of the setup.
The radiation reflected from the beam axis by the hole mirror
was extracted to the air through a KRS-5 window of 5 mm thickness.
First, it was focused by a parabolic mirror.
Three filters, a band-pass filter (BPF) and two ZnSe plates,  were located at the focal point.
The BPF removed the radiation contributed by mechanism other than the FEL.
The center wavelength and bandwidth of the BPF were designed to be 20 $\mu$m and 10\%.
The ZnSe plates were used as attenuators for the 20 $\mu$m radiation to avoid detector saturation.
The radiation was then again focused by a parabolic mirror,
and it was detected by a HgCdTe detector and a pyroelectric sensor.
The two detectors were mounted on a common three-axis stage.

%%%%%%%%%%%%%%
%%%%%%%%%%

The radiation output port downstream of the second undulator 
was prepared for the final FEL output.
A radiation measurement system 
and an application system were prepared on the table.
Figure \ref{fig:udetector_setup}(b) shows the layout of the setup.
The basic design was the same as the upstream design.
The first mirror switched the path to either of the two setups.
The reflection path was for a measurement system similar to that of the first undulator.
The transmission path (in the case without the switching mirror) was prepared 
for sending the radiation to the irradiation system.
Figure \ref{fig:udetector_setup}(c) shows the table for the second undulator.

To align the optical setup at the two systems,
a guide laser was prepared.
A He-Ne laser 
was placed upstream of the first undulator.
The guide laser
was placed on the beam axis through the undulators,
as shown in Fig. \ref{fig:undsection}.
If the hole mirrors at the FEL output ports were inserted 
with some offset for shifting the hole from the axis,
the guide laser was extracted to the radiation measurement systems.
The initial optical alignment of the measurement systems was performed with this guide laser.

The HgCdTe detector (FTIR-24-1.0)
is a product of InfraRed Associates.
It was used at the temperature of  liquid nitrogen and
was assembled with a dewar for 12~h operation.
A dedicated preamplifier was placed near the detector.
The voltage pulse signal was then recorded with an oscilloscope.
The sensitive wavelength range was between 7 and 24 $\mu$m with the KRS-5 window model.
The maximum sensitivity was at a wavelength of 15 $\mu$m.
The sensor size of the HgCdTe detector is 1~mm $\times$ 1~mm.
By scanning the detector in the horizontal and vertical directions,
the spatial profile of the radiation can be obtained.
By repeating the measurement while changing the longitudinal position of the detector,
we can set the detectors at the focal position.

The pyroelectric detector (QE8SP-B-MT-D0, M-Link)
is a product of Gentec Inc. 
The sensor was connected to its amplifier,
and the amplified analog output was recorded with an oscilloscope.
Because the time constant of the detector was much longer than the beam macro-pulse duration,
the measured signal corresponds to the total energy of the macro-pulse.
The sensor size of the pyroelectric detector is 7.8~mm $\times$ 7.8~mm.
After confirming that the spot size was sufficiently small compared with the sensor size,
we switched the detector from the HgCdTe to the pyroelectric sensor
for the absolute energy measurement.
Calibration of the pyroelectric detector was performed before the beam experiment.
Using an infrared (976~nm wavelength) laser of known pulse energy and similar pulse shape as the beam,
the output voltage of the amplifier could be calibrated
taking into account the spectrum difference in the absorption constant of the sensor material.

%----------------------------------------------------------
\subsection{Application demonstration setup}
%----------------------------------------------------------

To optimize the spot size and divergence of the FEL radiation
for the irradiation setup,
a telescope system was prepared.
The FEL output from the second undulator was
first focused with a parabolic mirror of 230 mm focal length,
and then converted to a collimated light 
with a parabolic mirror of 100 mm focal length at the over-focus position.
The distance between the parabolic mirror pair
can be controlled remotely
to adjust the divergence  downstream.
A switch mirror downstream of the telescope system
reflects the radiation to the irradiation system,
while the radiation is sent to the measurement system
of the HgCdTe detector on a three-axis stage
if the switch mirror is off the line.
The spatial profile and the divergence can be measured
by scanning the detector.

\begin{figure}[htb]
	\begin{center}
 	 \includegraphics[bb=0 0 700 550, width= 0.8\linewidth]{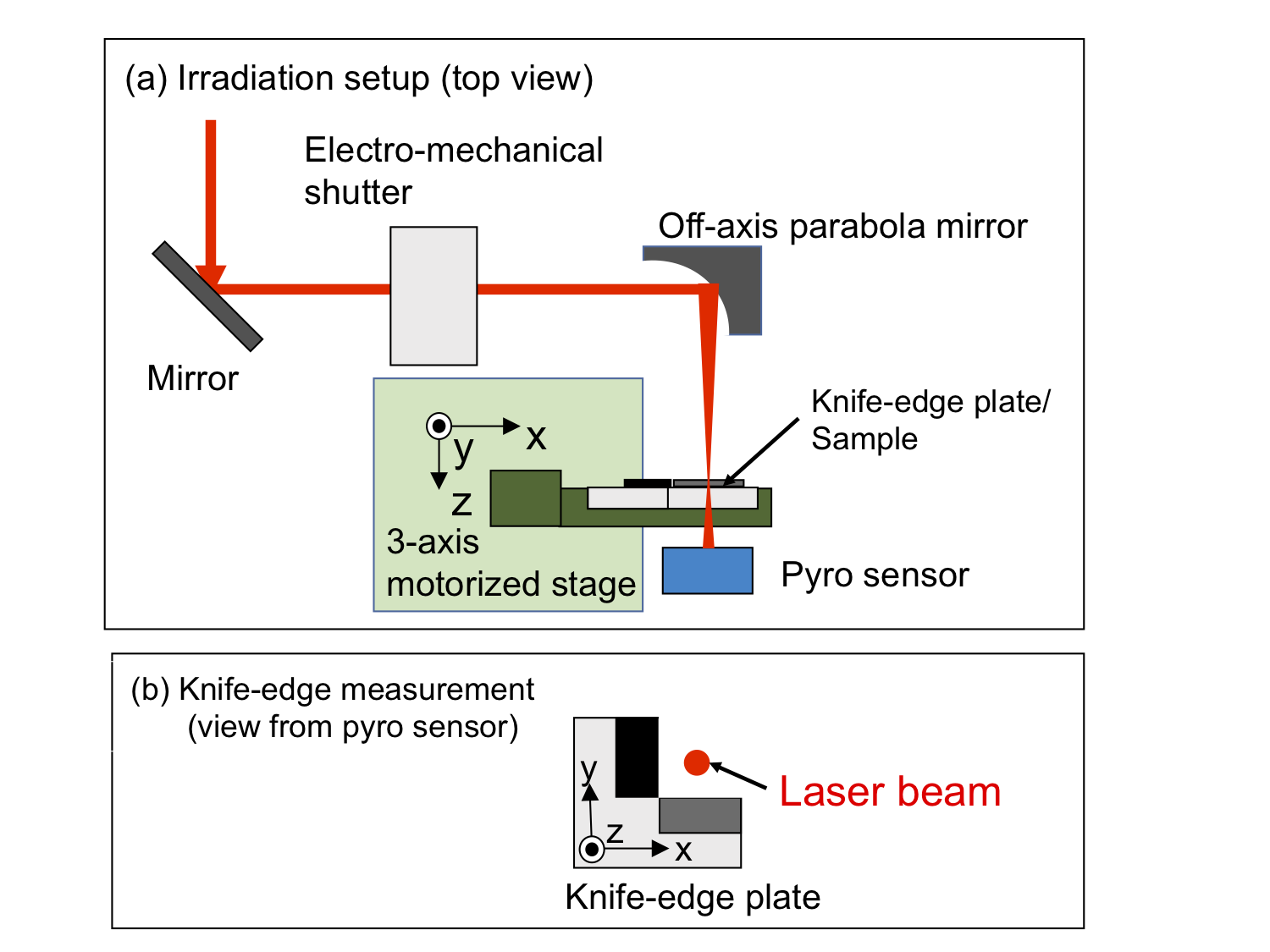}
	  \caption{
	  Setup for the irradiation experiment.
	  The FEL radiation is tightly focused on the sample by the parabola mirror.
	  The sample can be moved with the three-axis stage.
	  In the knife-edge beam size measurement,
	  the sample was replaced with a plate with mutually perpendicular knife edges.
}
	\label{fig:irrsystem}
\end{center}
\end{figure}

To investigate the effect of irradiation of mid-infrared FEL pulse, and the nonlinear response of materials 
to high-peak-intensity mid-infrared radiation,
the FEL irradiation system was developed.
Figure \ref{fig:irrsystem} shows the setup.
It consists of a parabolic mirror of 25.4~mm diameter and 50.8~mm focal length,
%（MPD129-M01, Thorlabs Inc.）,
to tightly focus the radiation on the sample
and a remote control stage system for setting the samples. 
Samples for the irradiation experiment, 
typically plates of 50~mm $\times$ 50~mm,
can be set on the stage.
A microscope system was prepared to monitor the sample in real time during the experiment.
A remote control shutter 
%(SH1, Thorlabs Inc) 
was prepared at the entrance of the system
to precisely control the irradiation time for the sample.

To confirm the spot size on the sample,
a knife-edge scan system \cite{skinner} was developed.
Two mutually perpendicular knife edge plates were mounted on the sample stage instead of the sample.
A pyroelectric detector 
%(QE8SP-B-MT-D0, M-Link, Gentec inc)
was set behind the sample stage.
%The beam radius (2sigma) are  plotted with z and fitted by an analytical Gaussian beam propagation equation to estimate beam quality factor M^2[ref M^2]. 

%%%%%%%%%%%%%%%%
%%%%%%%%%%%%%%%%

%===========================================
\section{EXPERIMENT}
\label{sec:experiment}
%===========================================

%---------------------------------------------------------------------------
\subsection{Beam Tuning}
%---------------------------------------------------------------------------
\subsubsection{Beam transport tuning and measured beam performance}
\label{sec beamtuning}

In this series of experiments,
the beam was operated in burst mode,
where all the components of the accelerator system were operated in continuous mode 
except for the laser system of the electron gun
 \cite{cerl_construction}.
This mode of operation was to enable beam operation allowing for beam loss.
The electron beam was emitted as a macro-pulse 
of 130 bunches with a 1.3~GHz fundamental repetition frequency.
The macro-pulse was emitted at 5~Hz.
The bunch charge and macro-pulse duration were confirmed at the Faraday cup 
located just downstream of the gun.

The upstream beam, starting from the gun cathode to the exit of the main linac,
was separately optimized by simulations.
Based on the simulation for the space-charge-dominated beam,
procedures to tune the beam from upstream were developed.
%Details of the upstream beam optimization are given in \cite{inj_tanaka}.
Here, we begin our description from the tuning of the second main linac cavity.

\begin{figure}[htb]
	\begin{center}
 	 \includegraphics[bb=0 0 600 400, width= 0.8\linewidth]{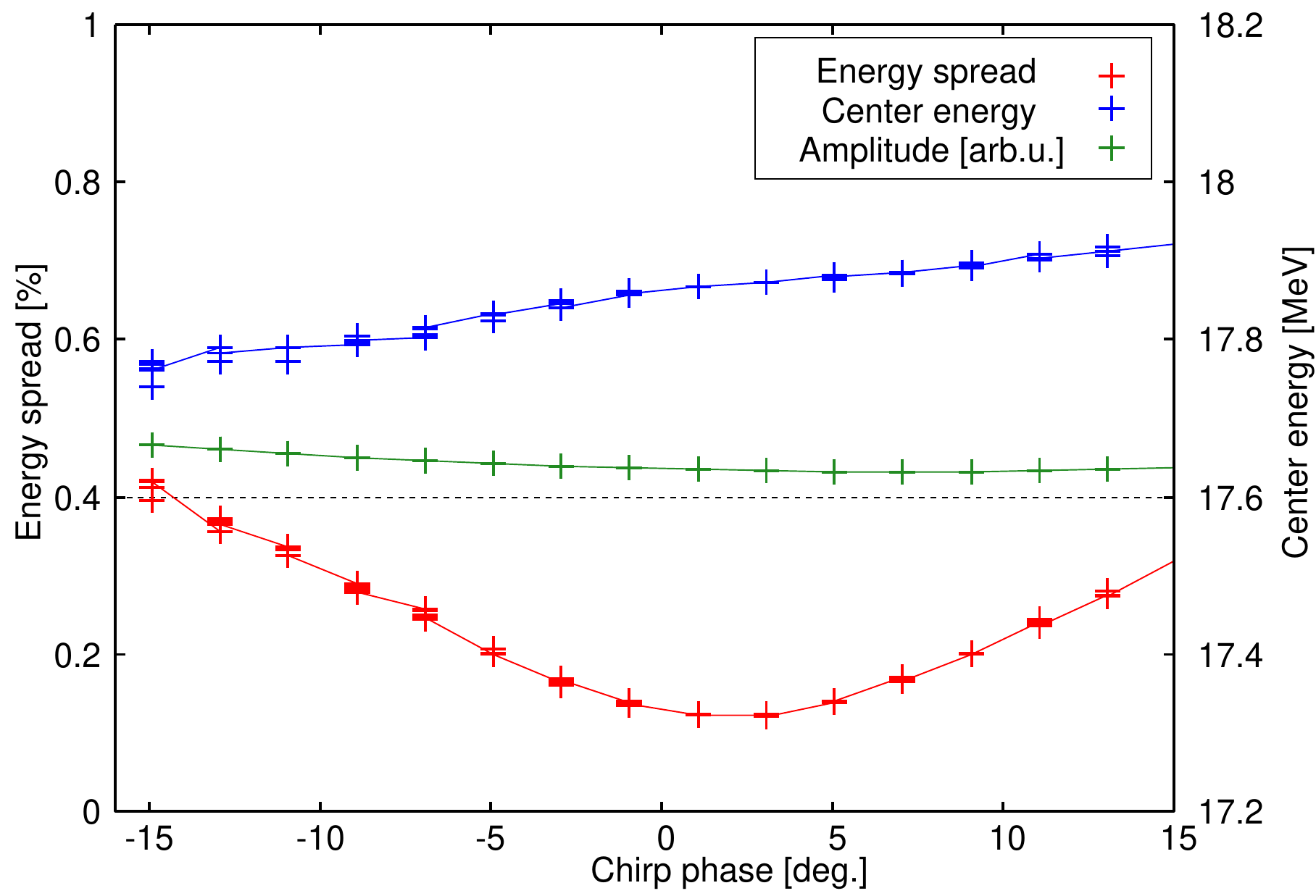}
	  \caption{
	  Measurement of the energy spread as a function of RF phase of the second main linac cavity.
	  The horizontal beam size and position at the profile monitor in the beginning of the arc section
	  were measured.
}
	\label{fig:ml2phasescan}
\end{center}
\end{figure}

To set the longitudinal energy chirp in the bunch,
we first surveyed the RF phase of the second main accelerator cavity.
Figure \ref{fig:ml2phasescan} shows the dependence of the horizontal beam size at a dispersive position in the arc section
on the RF phase while the beam energy is retained by simultaneous adjustment of the RF amplitude.
The minimum condition of the horizontal beam size
corresponds to the phase where the energy chirp is almost zero 
(according to the simulation, it corresponds to $+2^\circ$ off from the minumum condition of the energy spread).
This condition defines the origin of the phase, called as the chirp phase, in our discussion.
We set the RF phase to the positive direction,
which corresponds to the case that the bunch head has higher energy than the tail.
%Here, we call the phase as chirp phase.
The chirp phase is an important parameter to control bunch compression.
After rough trials to survey the starting condition before fine beam tuning for maximizing FEL,
we set our starting point of the chirp phase.

%%%%%%%%%%%%%%%%
%%%%%%%%%%%%%%%%

In the following, we show the measured beam parameters
obtained after  FEL maximization tuning, which will be explained
in Sec. \ref{sec:felmax}.
The results are summarized in Table \ref{tab:beamparam}.

%%%%%%%%%%%%%%%%%%%%%
\begin{table}[htb]
  \begin{center}
    \caption{Measured beam parameters after FEL optimization}
    \label{tab:beamparam}
    \begin{tabular}{l c | l } \hline
      Bunch charge & & 60 pC/bunch\\  \hline
      Exit of Main linac && \\ \hline
      Norm. Emittance (X) & $\epsilon_x$& 3.27$\pm$0.17~$\mu$m\\
      Norm. Emittance (Y) & $\epsilon_y$& 1.57$\pm$0.05~$\mu$m\\
      Energy spread & $\Delta \gamma/\gamma$& 0.6\%  \\
      Total Energy & $\sigma_t$& 17.7 MeV \\ \hline
      Transport && \\ \hline
      Chirp phase& &  +29$^\circ$ \\
      $R_{56}$ & & -0.282$\pm$0.046~m  \\ \hline
      Entrance of Undulator && \\ \hline
      Norm. Emittance (X) & $\epsilon_x$& 5.55$\pm$0.19~$\mu$m\\
      Norm. Emittance (Y) & $\epsilon_y$& 5.11$\pm$0.16~$\mu$m\\
      Bunch length (RMS) & $\sigma_t$& $\sim$ps (0.6~ps structure) \\   
      \hline
  \end{tabular}
  \end{center}
\end{table}
%%%%%%%%%%%%%%%%%%%%%%

%Figure \ref{fig:energyspread} shows the energy distribution
The energy spread was
measured by the horizontal beam profile at a dispersive position in the arc section.
At the profile monitor, 
the horizontal dispersion and beta function were calculated as $\eta_x=0.49$~m and $\beta_x=1.9$~m, respectively.
The horizontal beam size was dominated by the dispersion term.
The RMS energy spread was estimated as 0.63\%.
Note that the energy spread
% in Fig.\ref{fig:energyspread} 
included contribution of the whole macro-pulse.
The beam loading effect of macro-pulse charge, 60~pC $\times$ 130~bunches,
can change the accelerating field in RF cavities along the bunch train.
By calculating the beam loading effect,
the beam energy change from pulse head to tail is estimated as 0.5\%.
In the energy spread measurement,
a contribution of 0.14\% in RMS was made by this pulse-to-pulse effect.

\begin{figure}[htb]
	\begin{center}
 	 \includegraphics[bb=0 0 500 320, width= 0.8\linewidth]{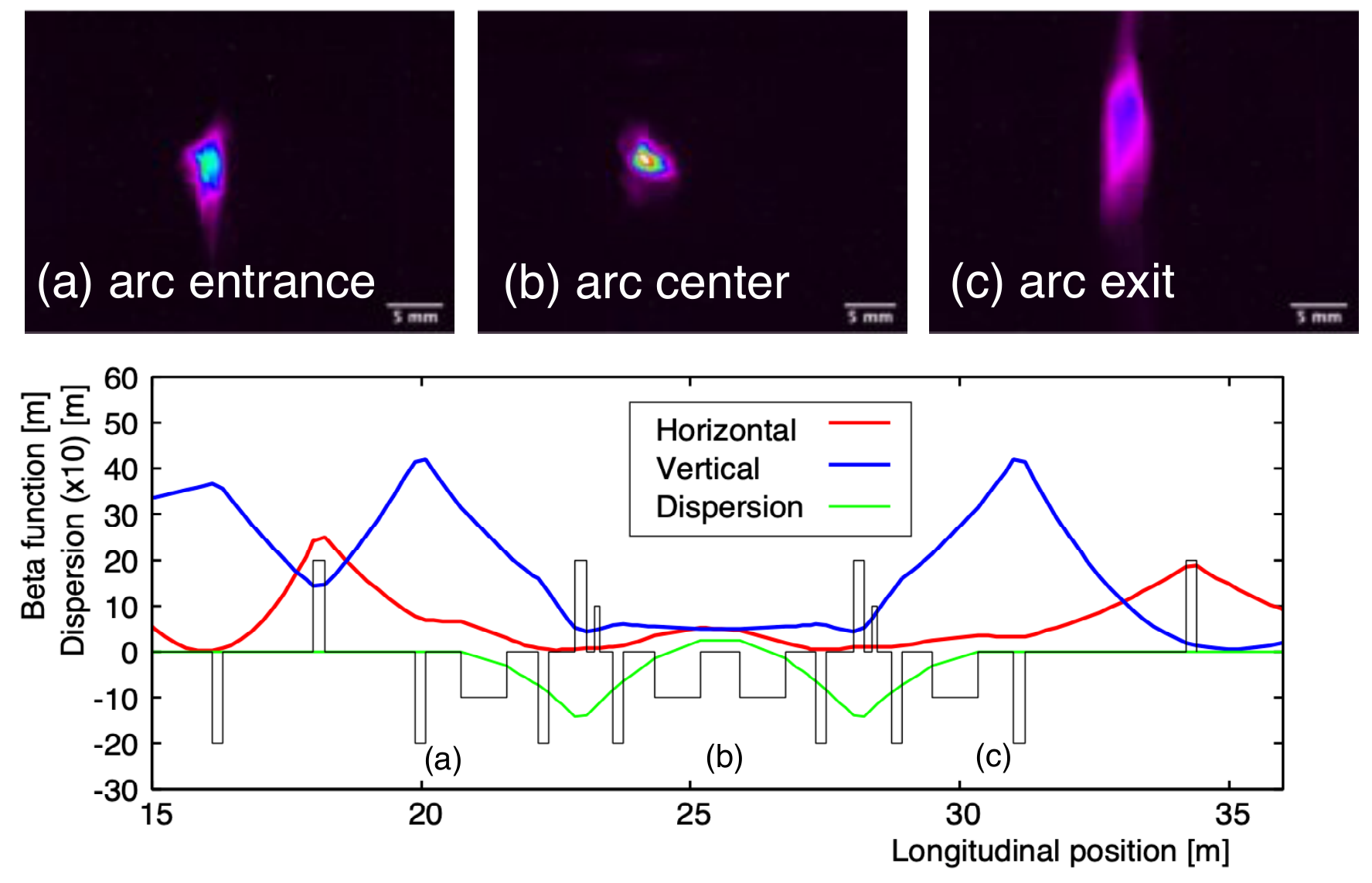}
	  \caption{
	  Beam profiles in the arc section.
	  The arc was designed to be symmetric with respect to the center.
}
	\label{fig:arcprofile}
\end{center}
\end{figure}

Because the arc section is designed to have a strong vertical focusing,
the beam profile and divergence after the arc are sensitive to the transverse beam parameters at the arc entrance.
The orthogonal combined knobs 
using four quadrupole magnets just before the arc were prepared
as explained in \ref{sec:knobs}.
Figure \ref{fig:arcprofile} shows the beam profile
measured at the arc entrance, at the center, and at the exit.
It shows that the symmetric optics with respect to the arc center
was successfully been realized.

%%%%%%%%%%%%%%%%
%%%%%%%%%%%%%%%%

\begin{figure}[htb]
	\begin{center}
 	 \includegraphics[bb=0 0 320 220, width= 0.8\linewidth]{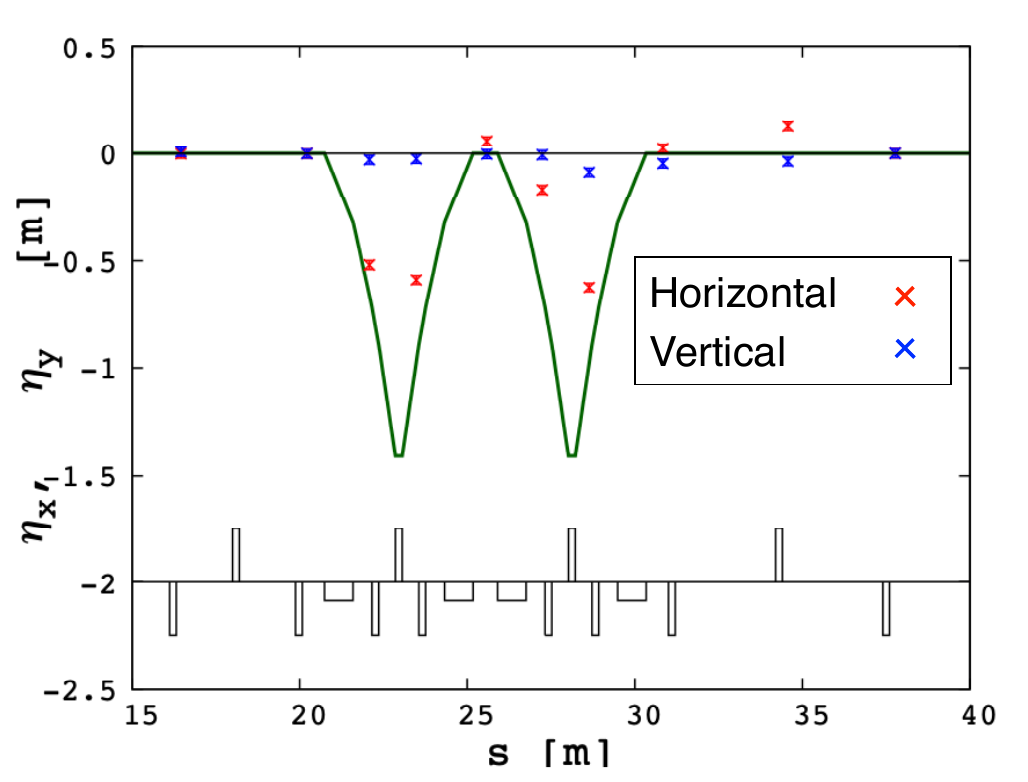}
	  \caption{
	  Dispersion measurement at the arc section.
	The beam energy dependence of the transverse beam orbit
	was measured by the beam position monitors.
}
	\label{fig:dispersionmeas}
\end{center}
\end{figure}

To confirm the achromat condition of the arc, the transverse dispersion was measured.
The RF amplitude of the second main linac cavity was changed to shift the beam energy by $-1$\%,
and the beam position change was measured by the beam position monitors.
In this measurement, sextupole magnets were set off 
because they introduce a strong nonlinear effect in this measurement of a relatively large orbit change.
Figure \ref{fig:dispersionmeas} shows the results of the measurements,
which confirmed that
the dispersion at downstream of the arc and coupling in the vertical direction were small.
We also confirmed that 
the achromat condition was retained 
while  the $R_{56}$ knobs were being changed.
From the symmetric design of the arc,
the transverse dispersion at the center of the arc $\eta_c$
and the $R_{56}$ of the arc are related by
$R_{56} = 1.414  \eta_c - 0.34 \quad [\textrm{m}]$,
 in the present case \cite{cerl_bunch}.
By measuring $\eta_c$,
the $R_{56}$ of the arc can be estimated indirectly.
%Another method to check $R_{56}$ is by using a time-of-flight (TOF) measurement \cite{cerl_shimada}.
%By measuring the bunch arrival time at the exit of the arc using the analog signal of a beam position monitor pick-up,
%the relative TOF of the arc was measured.
%By measuring the TOF while changing the beam energy,
%$R_{56}$ could be directly estimated.
The $R_{56}$ at the FEL power maximum was
estimated as $-0.282\pm0.046 $~m.

%%%%%%%%%%%%%%%%
%%%%%%%%%%%%%%%%

\begin{figure}[htb]
	\begin{center}
 	 \includegraphics[bb=0 0 600 400, width= 0.8\linewidth]{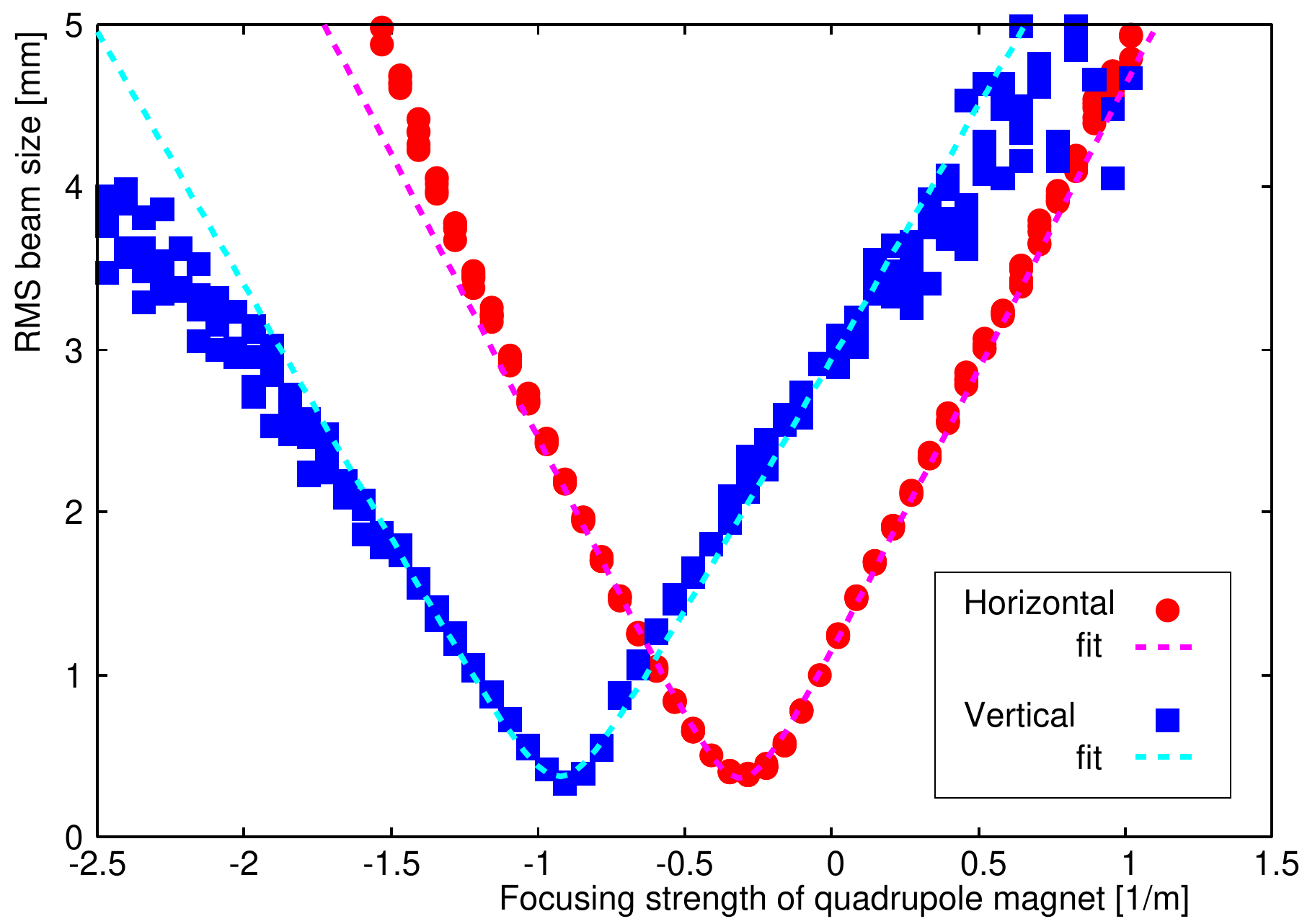}
	  \caption{
	  Measurement of emittance at the upstream of the FEL section.
	  The beam size at the screen monitor was measured
	  while the strength of a quadrupole magnet was being scanned.
	  The line is the fitting result obtained by using a function $\sqrt{a (K-b)^2 + c}$,
	  leaving $a,b,$ and $c$ as free parameters.
}
	\label{fig:emittancemeas}
\end{center}
\end{figure}

The beam emittance just upstream of the FEL section
was measured by the waist scan method using a quadrupole magnet and a profile monitor
\cite{cerl_akagi}.
Figure \ref{fig:emittancemeas} shows the results.
By fitting the data with a hyperbolic function,
the normalized emittance was estimated as
$\epsilon_x=5.55\pm0.19~\mu$m and $\epsilon_y=5.11\pm0.16~\mu$m
for the horizontal and vertical directions, respectively.
The emittance measured at just downstream of the main linac
using a similar method
resulted in 
$\epsilon_x=3.27 \pm 0.17~\mu$m and $\epsilon_y=1.57 \pm0.05~\mu$m
for the horizontal and vertical directions, respectively.
There was some emittance degradation in the beam transport,
but, the effect was still within the manageable range.

%%%%%%%%%%%%%%%%
%%%%%%%%%%%%%%%%

\begin{figure}[htb]
	\begin{center}
 	 \includegraphics[bb=0 0 800 400, width= 0.9\linewidth]{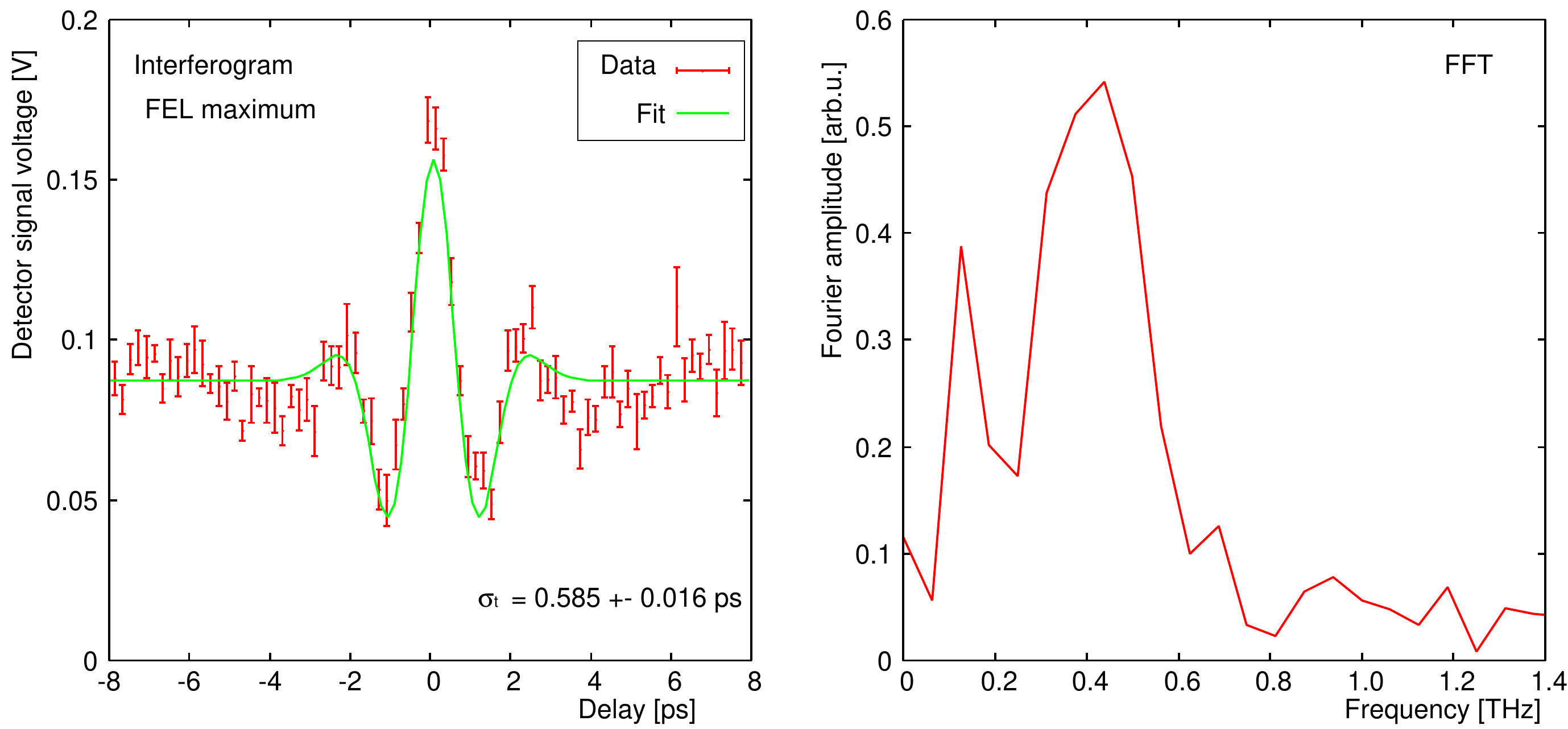}
	  \caption{
	  Autocorrelation signal of CTR interferometer (left) and its spectrum (right).
	  By fitting the data with a model function of low-frequency cutoff,
	  the RMS bunch length of the core part of the bunch
	  can be estimated indirectly.
}
	\label{fig:ctrinterferometer}
\end{center}
\end{figure}

The bunch length is an important parameter for the FEL.
However, the existing setup did not have an instrument to measure the bunch length directly.
We used coherent radiation as an indirect tool to evaluate the bunch length.
A terahertz interferometer system using coherent transition radiation (CTR)
was developed
approximately 10~m upstream of the FEL undulator in the same straight section
\cite{cerl_bunch}.
By detecting the autocorrelation with a Michelson interferometer,
the spectrum of the bunch can be obtained.
Figure \ref{fig:ctrinterferometer} shows the interferogram data
and the results of the fitting analysis.
It shows an RMS bunch length of $\sigma_t=0.585 \pm 0.016$ ps.
To check the bunch length measurement in a relatively direct way,
we tried measuring incoherent transition radiation in the visible range using a streak camera system.
Although the measurement was performed in a different beam operation period, 
and hence the details of beam tuning were not exactly the same,
the typical beam parameters were similar.
The result was never shorter than 1.6 ps in RMS.
Comparing the results obtained with the streak camera with those obtained with the interferometer,
we infer that the scale of the whole bunch size was in the range of a few picoseconds,
and it had small longitudinal structures of 0.6 ps scale at the same time.

%%%%%%%%%%%%%%%%%
%%%%%%%%%%%%%%%%%

\begin{figure}[htb]
	\begin{center}
 	 \includegraphics[bb=0 0 650 500, width= 0.8\linewidth]{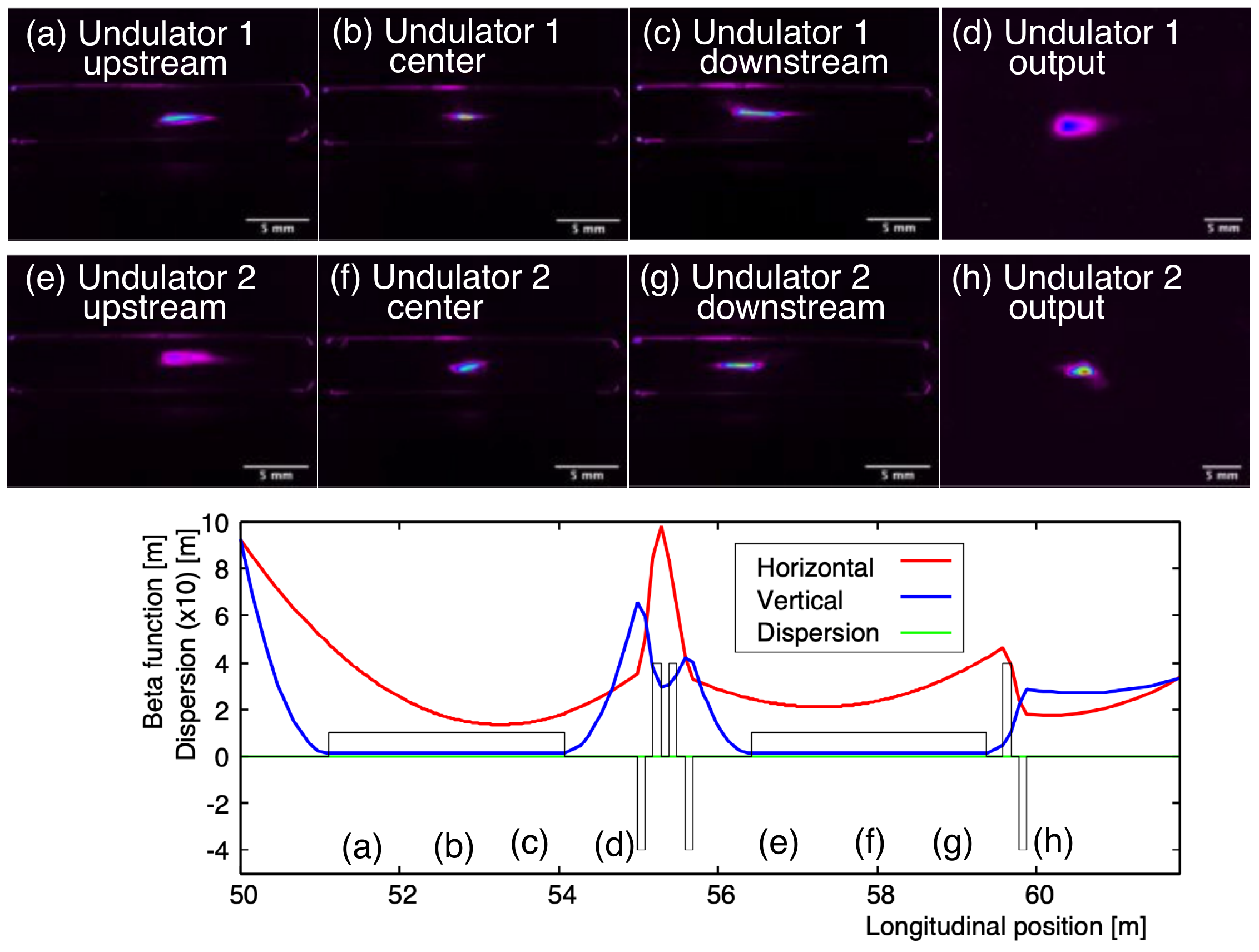}
	  \caption{
	  Beam profile at the FEL section.
	  Three screen monitors in each undulator duct were used to confirm beam optics matching.
	  The two FEL output ports at downstream of each undulators also have screen monitors.
}
	\label{fig:undulatorscreen}
\end{center}
\end{figure}

The transverse beam matching to the undulator was performed
using
four orthogonal knobs 
while measuring the screen monitor in the undulator duct.
Figure \ref{fig:undulatorscreen} shows a typical result after the tuning. 
It shows that the vertical beam size remained constant
as the vertical Twiss parameters matched to the undulator focusing,
and that the horizontal beam size become minimum at the center of each undulator
to minimize the averaged beam size in the undulator.

%%%%%%%%%%%%%%%%%
%%%%%%%%%%%%%%%%%

\subsubsection{FEL power maximization tuning}
\label{sec:felmax}

We explain the beam tuning procedure for FEL power maximization in the following.
Once the beam condition has been established to pass the undulators with a reasonable beam size and position,
the tuning procedure was switched to the scheme based on the FEL radiation measurements.

\begin{figure}[htb]
	\begin{center}
 	 \includegraphics[bb=0 0 500 250, width= 0.8\linewidth]{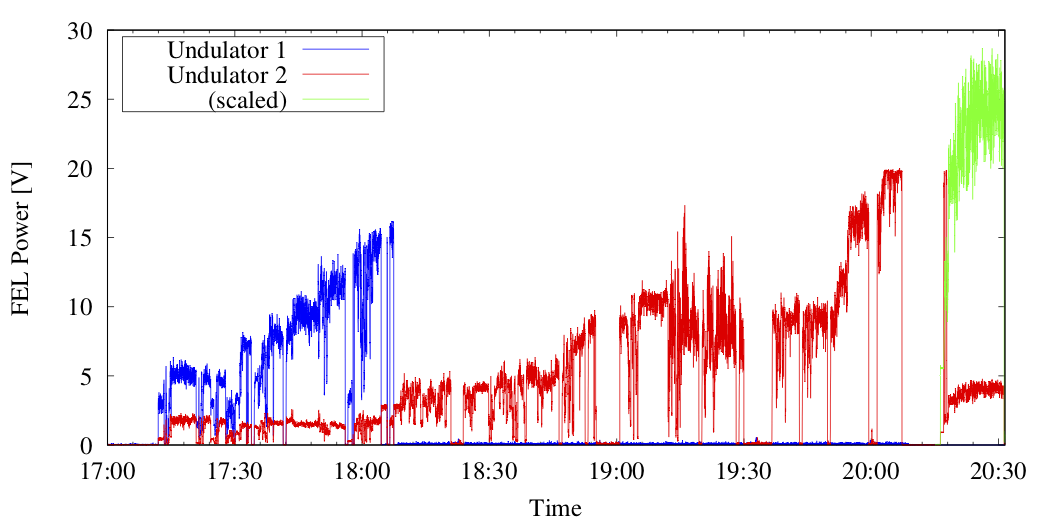}
	  \caption{
	  Trend of FEL power during the automatic tuning.
	  Because the detector reached saturation (at 20:00),
	  we changed the detector position longitudinally and continued the tuning.
	  To correct the discontinuity of the detector signal, a scaled data (after 20:15) is also shown.
	  The final absolute energy measurement was performed 
	  by switching to the pyrodetector.
}
	\label{fig:tuning_trend}
\end{center}
\end{figure}

To perform
the beam tuning efficiently,
we have prepared several knobs for use in the beam operation
among the many control parameters in the complicated accelerator system.
The knobs routinely used were as follows:
\begin{itemize}
\item
The $R_{56}$ knob changes the quadrupole magnets in the arc.
Because $R_{56}$ changes the longitudinal distribution of the bunch, it is a sensitive knob for the FEL.
\item
The SX knobs were the two sextupole magnets in the arc section.
They change the higher-order terms of the transverse dispersion, 
as a result of which the shape of the longitudinal phase space distribution changes nonlinearly.
\item
Matching knobs for the first and second undulator are 
the orthogonal combined knobs with four quadrupole magnets.
They change the average beam size along the undulator,
and should affect the FEL gain.
\item
The injector RF parameters, namely, the phase and amplitude of the buncher and the first cavity in the injector module.
Because the velocity bunching at the low-energy section is sensitive to the initial bunch length,
they are sensitive parameters for the FEL.
\item
The RF phase and amplitude of the second cavity in the main linac.
They shift the chirp phase from the one initially chosen only after a rough survey.
\item
Steering dipole magnets at the entrance of each undulator.
They change the position and angle of the incoming beam orbit in both planes.
\end{itemize}
We applied a machine learning method based on a Gaussian process optimization \cite{gaussianprocess}.
A Python code using the GPyOpt library \cite{gpy} was used in the operation.
The evaluation parameter was the radiation intensity measured by HgCdTe detectors.
The parameters to change were the knobs listed earlier.
Typically three or four knobs were used simultaneously in the optimization,
and the best parameter set was found after approximately 50 epochs.
Then the optimization continued using other knobs.
In the first half,
the optimization maximized the first undulator output,
after which it was switched to the second undulator output.
Figure \ref{fig:tuning_trend} shows an example of the FEL power trend
during the FEL optimization.
Typically, it took approximately 4~h to establish maximum power at the second undulator.
%%%%%%%%%%%%%%%
%%%%%%%%%%%%%%%

After the FEL maximization tuning,
we measured the chirp phase.
Since the auto tuning procedure,
especially the knobs of injector RF and main linac RF,
can change the longitudinal energy chirp of the bunch.
The final chirp phase at maximized FEL was measured to be +29$^\circ$ for  1.3 GHz RF.

\begin{figure}[htb]
	\begin{center}
 	 \includegraphics[bb=0 0 600 400, width= 0.8\linewidth]{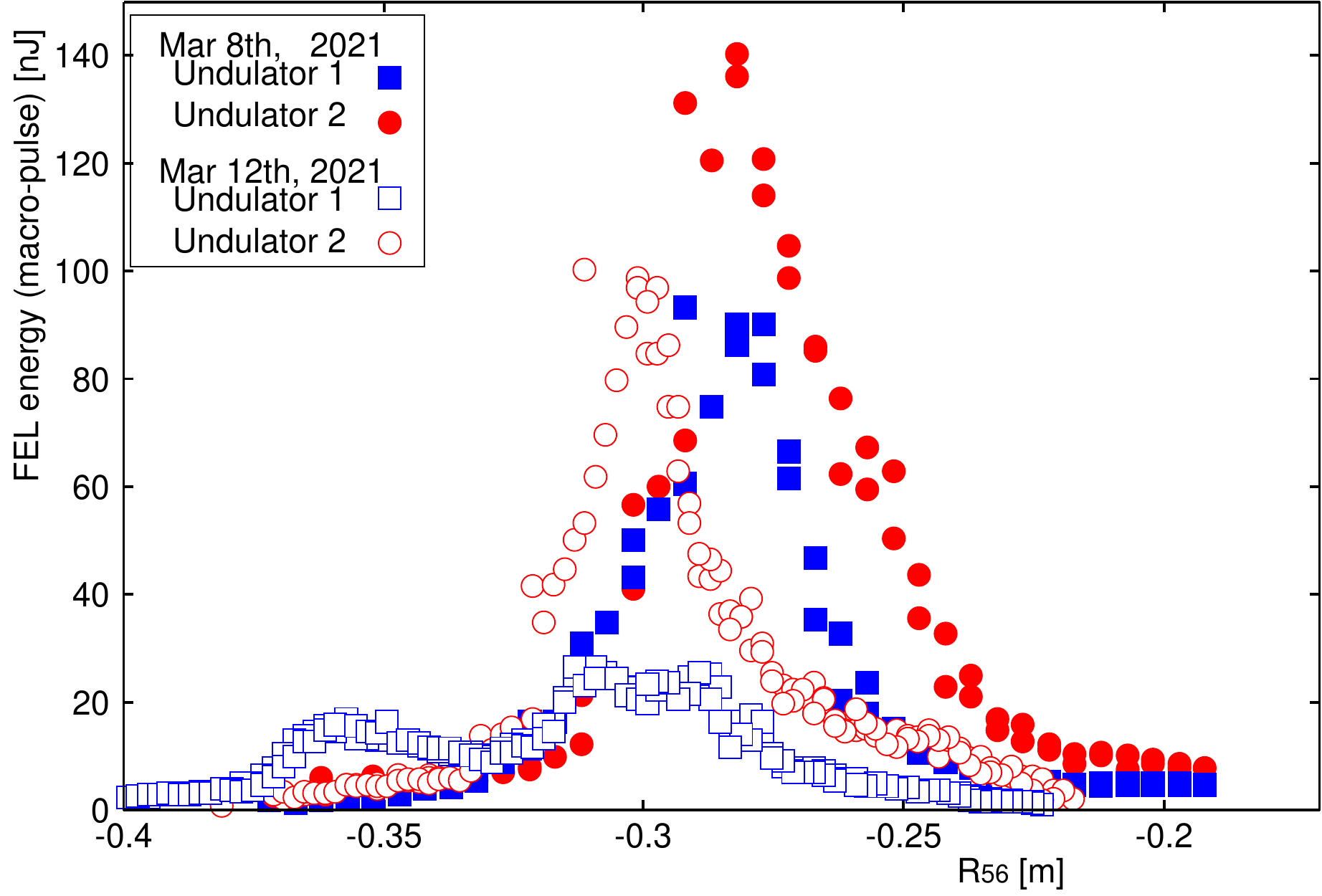}
	  \caption{
	  Response of FEL output power to the $R_{56}$ knob.
	  Two sets of measurements conducted on different days are shown.
	  The FEL tuning procedure was repeated at the start of the beam operation on each day.
}
	\label{fig:r56response}
\end{center}
\end{figure}

After we tuned the FEL output using auto-optimization,
we measured the final beam parameters described in Sec. \ref{sec beamtuning}
and the responses of the knobs over a wide range to check their sensitivity.
Figure \ref{fig:r56response} shows the FEL output response to the $R_{56}$ knob.
The FEL output of both undulators became maximum under the same condition.
The FWHMs of the peaks were approximately  0.03~m in $R_{56}$.

%%%%%%%%%%%%%%%
%%%%%%%%%%%%%%%

\begin{figure}[htb]
	\begin{center}
 	 \includegraphics[bb=0 0 600 400, width= 0.8\linewidth]{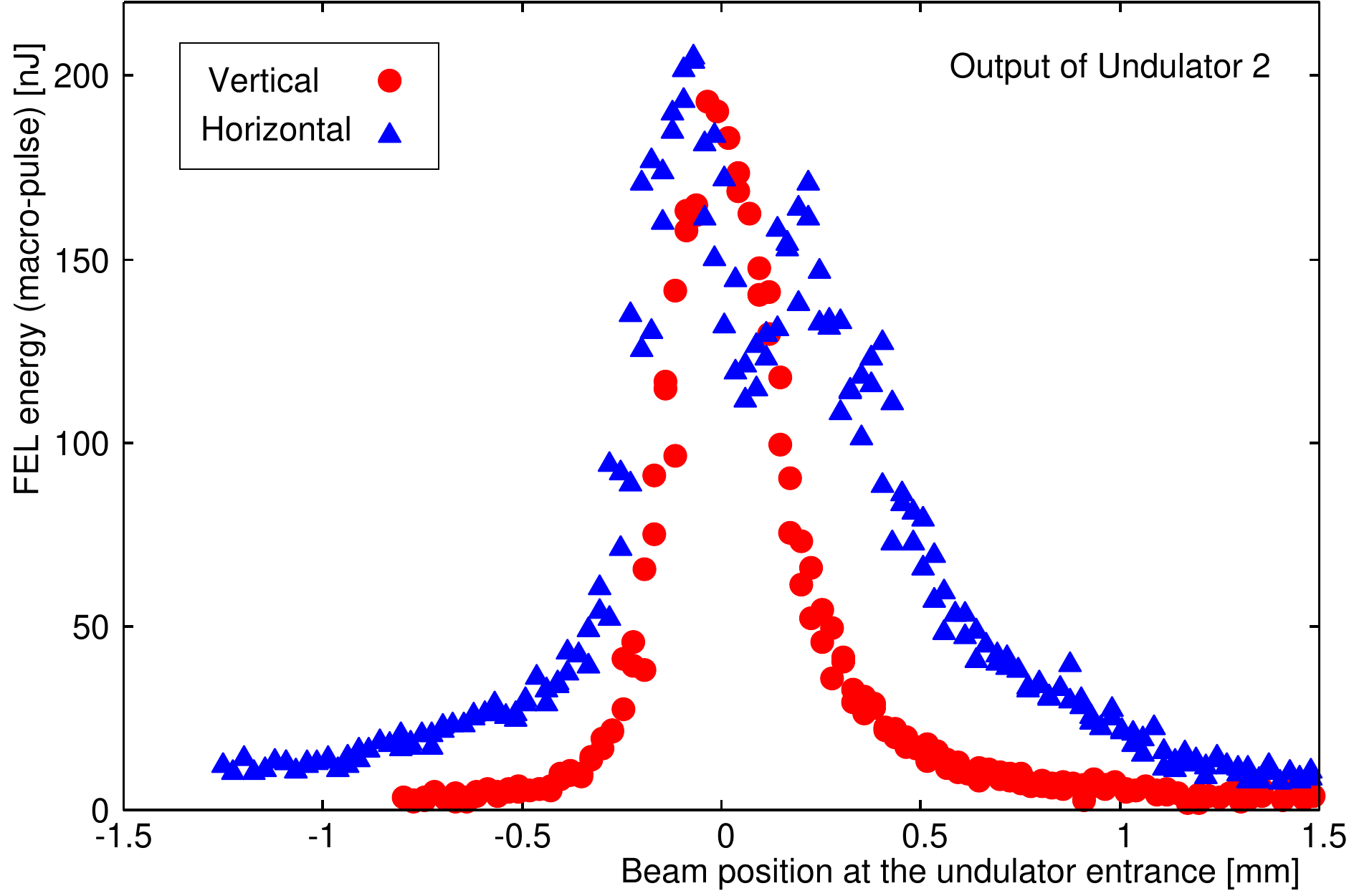}
	  \caption{
	  Response of FEL output power to the upstream steering magnet.
	  Two sets of measurements, vertical and horizontal scans,  are shown.
	  The horizontal axis is the estimated beam position change at the entrance of the first undulator.
}
	\label{fig:steeringresponse}
\end{center}
\end{figure}

Figure \ref{fig:steeringresponse} shows the FEL output response to the
vertical and horizontal steering magnets  upstream of the first undulator.
The FWHMs of the peaks were approximately 0.4~mm (vertical) and 0.8~mm (horizontal) of the beam position offset at the undulator entrance.

%%%%%%%%%%%%%%%
%%%%%%%%%%%%%%%

%--------------------------------------------------------------
\subsection{FEL characterization}
%--------------------------------------------------------------

\begin{figure}[htb]
	\begin{center}
 	 \includegraphics[bb=0 0 600 600, width= 0.8\linewidth]{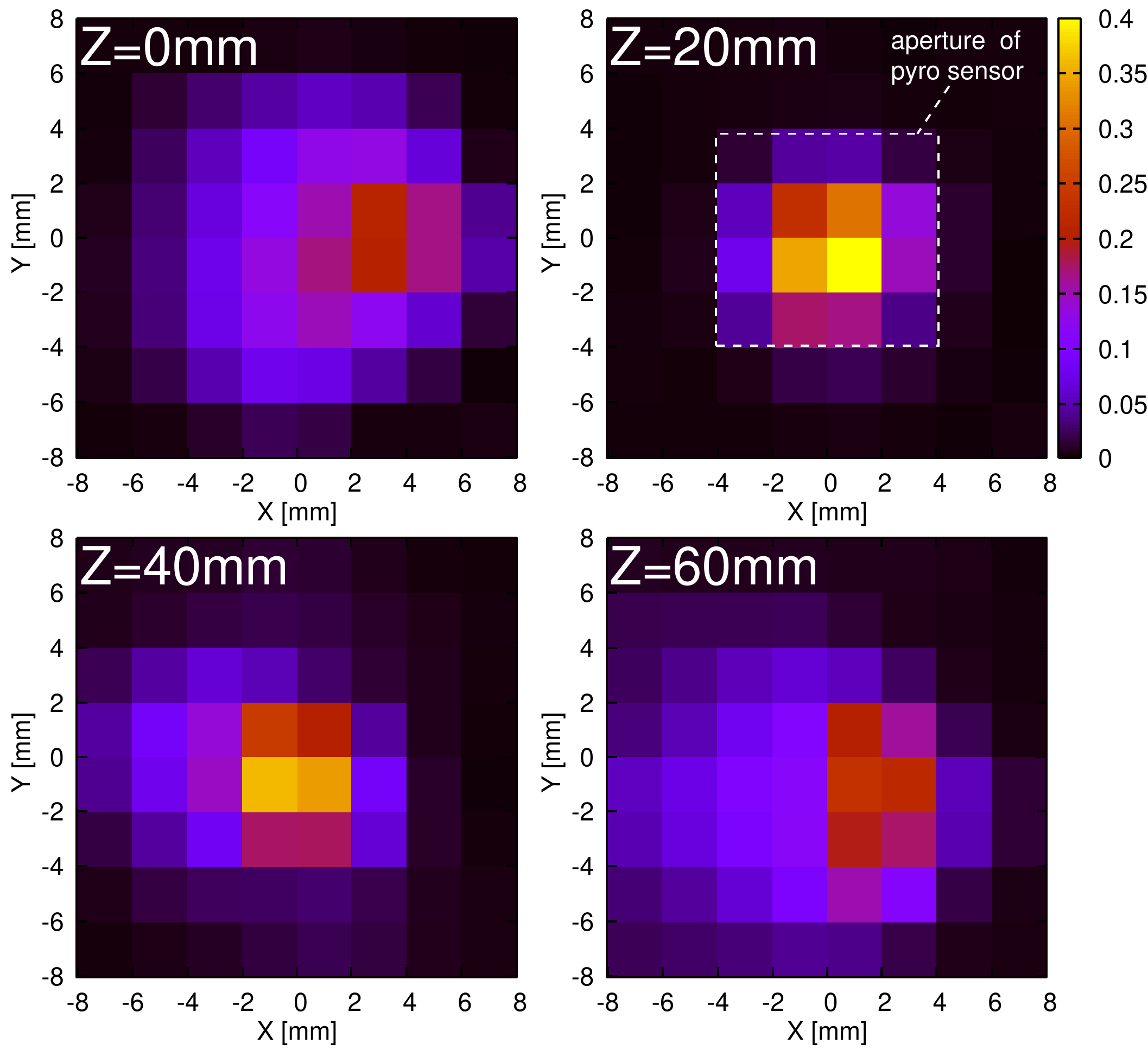}
	  \caption{
	  FEL spatial profiles at the second undulator output scanned around the focal point of the parabolic mirror.
	$Z$ denotes the longitudinal position of the detector.
	It was confirmed that the spot size was minimum at $Z=20$~mm
	and was smaller than the aperture of the pyrosensor used for total energy measurement.
}
	\label{fig:profileU2}
\end{center}
\end{figure}

Figure \ref{fig:profileU2} shows the spatial profile of radiation at the second undulator
measured by scanning the HgCdTe detector in two dimensions.
%Note that because the profile measurement system was placed downstream of the parabolic mirrors
%as shown in Fig. \ref{fig:udetector_setup},
%and the result was not the direct size and divergence from the source.
By changing the longitudinal position,
the spot size change around the focal position could be measured.
The pyroelectric detector measurements were performed at the focal position.

%%%%%%%%%%%%%%%%%
%%%%%%%%%%%%%%%%%

\begin{figure}[htb]
	\begin{center}
 	 \includegraphics[bb=0 0 600 400, width= 0.7\linewidth]{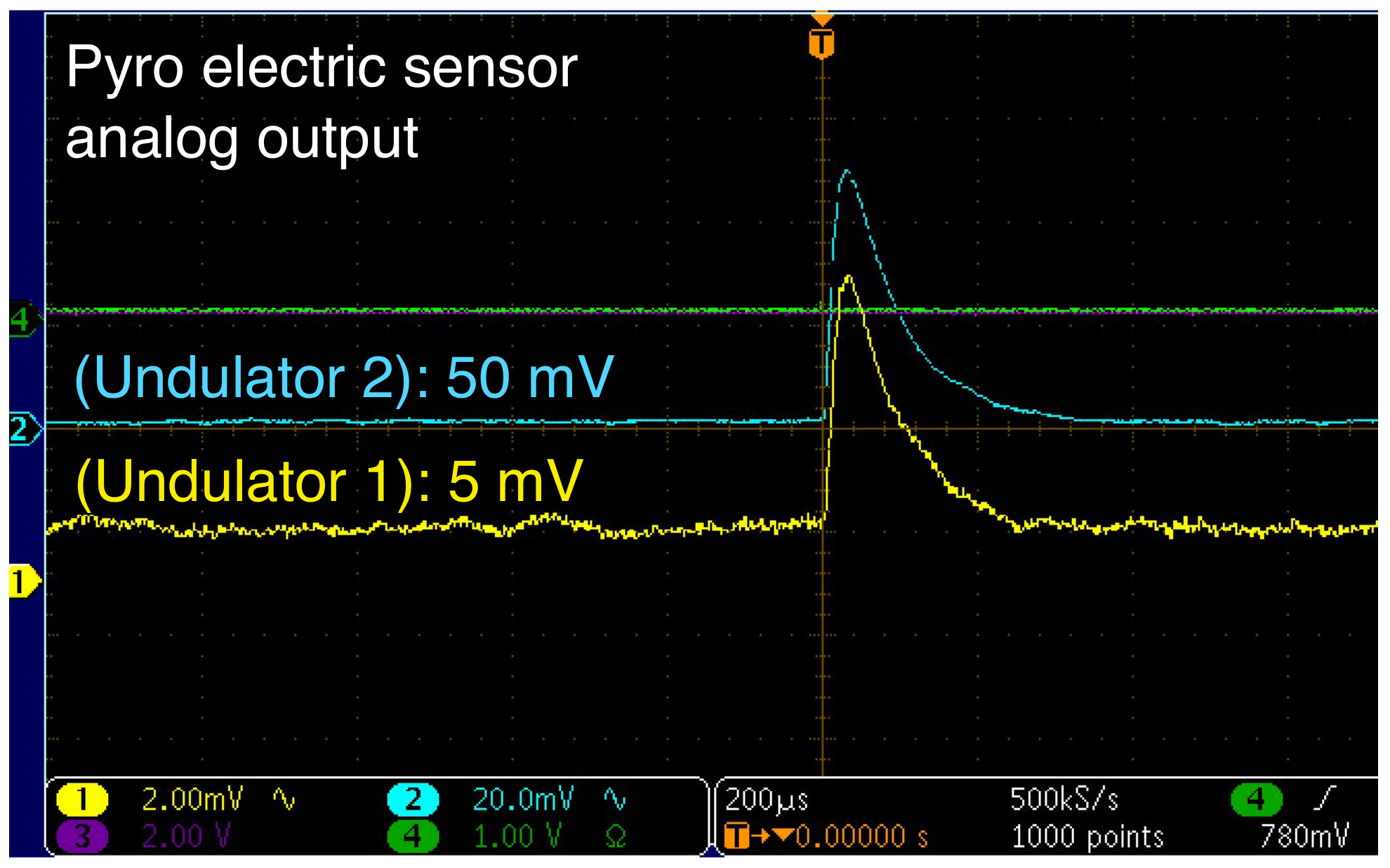}
	  \caption{
	  Absolute energy measurement of FEL macro-pulse. 
	  Analog signals of the amplifier of pyroelectric detectors were recorded.
	  The peak voltage can be converted to the absolute energy 
	  using a calibration coefficient of 6.9~nJ/mV at 20~$\mu$m wavelength.
}
	\label{fig:energymeter}
\end{center}
\end{figure}

\indent
After the FEL optimization tuning,
the absolute energy of the radiation was measured with the calibrated pyroelectric detector.
Figure \ref{fig:energymeter} shows the analog signal data when the maximum power was recorded.
The BPF was inserted to measure the 20~$\mu$m component.
The detectors were calibrated as 6.0~nJ/mV at an infrared wavelength of 976~nm,
and it becomes 6.9~nJ/mV at 20~$\mu$m according to the absorption spectrum.
Using this calibration,
the energies directly measured at each detector were 34.5~nJ/macro-pulse and 345~nJ/macro-pulse
for the first and second output port, respectively.
To evaluate the original energy at the undulator in vacuum,
the following corrections should be applied:
The reflectance of gold-plated mirrors--reflectance of each mirror was estimated as 0.95.
The transmission of the BPF at a wavelength of 20~$\mu$m was measured as 0.80.
The transmission of the KRS-5 window was measured as 0.78.
The geometrical efficiencies of hole mirror were estimated as 0.72 (first undulator) and 0.95 (second undulator),
which were based on the simulation of the FEL radiation spatial profile.
Including all these corrections,
the energies at each undulator were 94~nJ/macro-pulse and 753~nJ/macro-pulse.
Because each macro-pulse consisted of 130 bunches,
it corresponded to 0.72~nJ/bunch and 5.8~nJ/bunch.

%%%%%%%%%%%%%%%
%%%%%%%%%%%%%%%

\begin{figure}[htb]
	\begin{center}
 	 \includegraphics[bb=0 0 600 400, width= 0.7\linewidth]{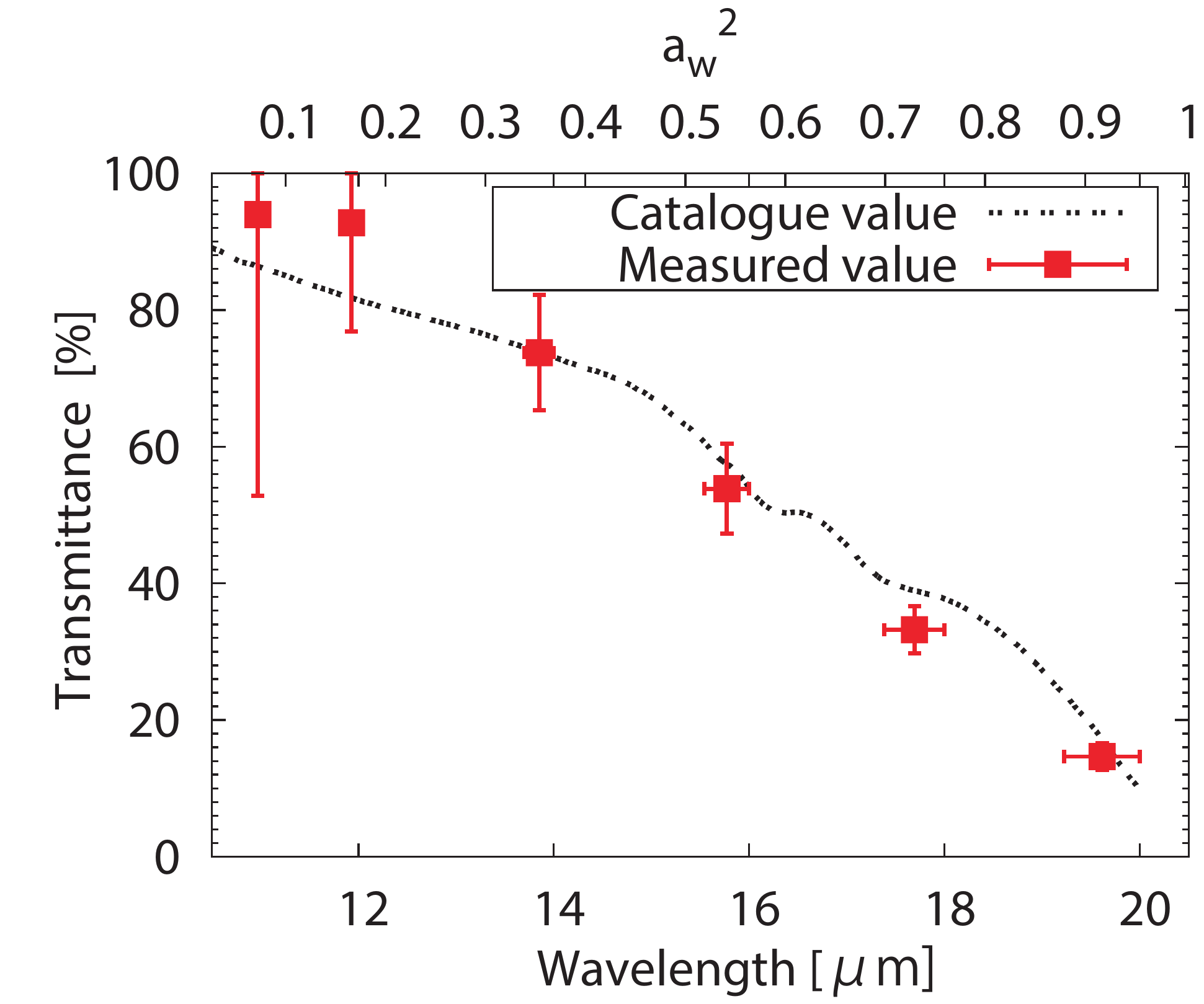}
	  \caption{
	  Transmission measurement of the ZnSe sample for showing wavelength tunability.
	  The signal ratio of the HgCdTe detector between the conditions with and without the sample were measured
	  while changing the undulator phase.
	  Assuming the calculated wavelength change,
	  the dependence was consistent with the transmission spectrum of the sample specified in the catalog.
}
	\label{fig:znsetransmission}
\end{center}
\end{figure}

Although a spectrometer system was not ready in this first commissioning,
we measured the center wavelength of the radiation using a simple method. 
The ZnSe plate (Thorlabs, WG71050-E3)
used as an attenuator was utilized as a standard sample that has a known transmission spectrum.
By measuring the HgCdTe signal with and without the ZnSe plate,
the transmission was measured.
The BPF was off in this measurement.
By repeating the measurement while changing the phase of both APU undulators simultaneously,
we obtained the transmission curve in Fig. \ref{fig:znsetransmission}.
The data were consistent with 
the specified transmission curve of the ZnSe plate (anti-reflective coating).

%%%%%%%%%%%%%%%
%%%%%%%%%%%%%%%

To show the stability of the FEL output,
we measured the radiation intensity in a free run of 30~min.
The measurement was performed with the HgCdTe detectors,
and their absolute energy was calibrated by comparing with the pyrosensor just before the measurement.
The shot-by-shot fluctuation was measured as 7\% in RMS
for both undulators.
Note that a single shot of macro-pulse consists of 130 bunches.
Assuming a non-correlated random Gaussian distribution,
the bunch-by-bunch fluctuation is estimated as 80\%.

%%%%%%%%%%%%%%%%%
\begin{figure}[htb]
	\begin{center}
 	 \includegraphics[bb=0 0 600 400, width= 0.8\linewidth]{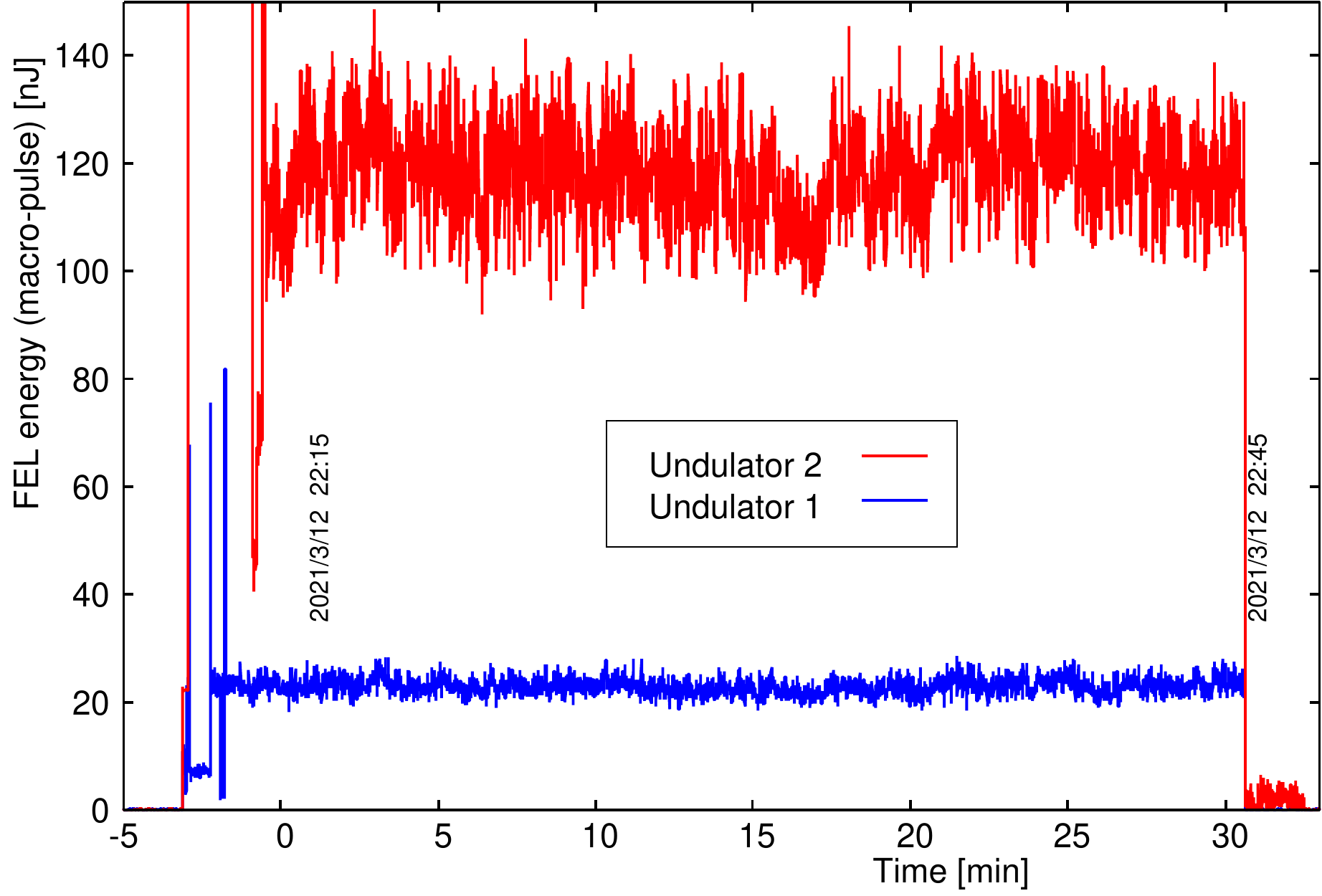}
	  \caption{
	  Trend of FEL energy in a free run of 30 min.
	  Shot-by-shot voltage of the detectors was recorded continuously.
	  (The beginning part of the data is for setting the detectors at non-saturated condition.)
}
	\label{fig:stability_trend}
\end{center}
\end{figure}
%%%%%%%%%%%%%%%%%

\subsection{Characterization of Focused Laser Beam for Application Experiments}

\begin{figure}[htb]
	\begin{center}
 	 \includegraphics[bb=0 0 350 350, width= 0.8\linewidth]{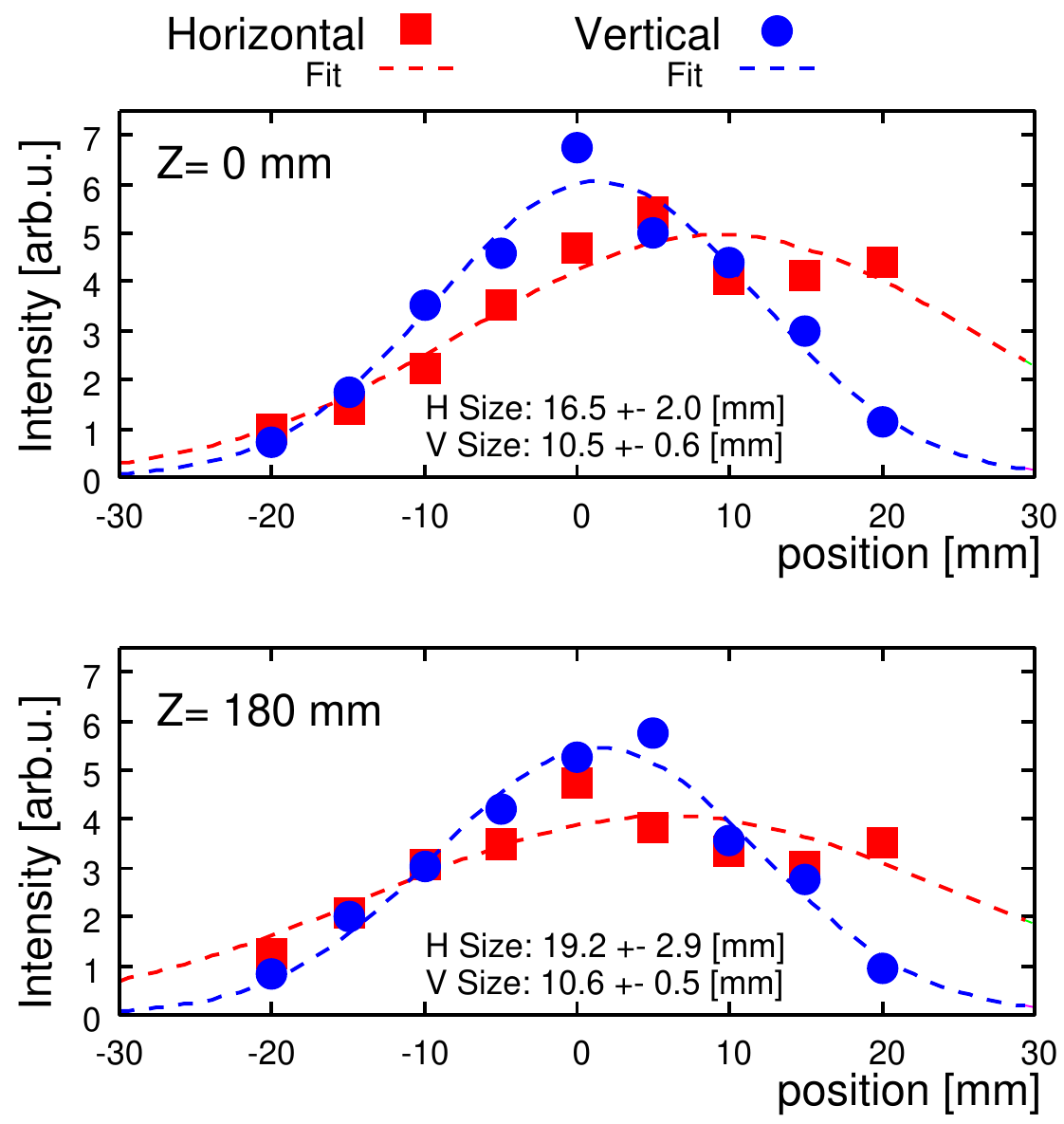}
	  \caption{
	  Results of FEL spatial profile measurements downstream of the telescope.
	  Measurements were performed at two longitudinal positions, 
	  namely, at $Z=0$ and 180~mm.
	  The projected distribution in two axes were fitted with a Gaussian function.
	  Because the sizes were the same in the two positions, 
	  the telescope was well adjusted for converting the FEL to a collimated beam.
}
	\label{fig:parallel}
\end{center}
\end{figure}

For the  FEL  application experiments,
we sent the radiation to the irradiation setup.
To confirm the spot size and its divergence,
spatial profile measurements were performed at downstream of the telescope system.
Figure \ref{fig:parallel} shows the results after the telescope tuning,
adjustment of the distance between the parabolic mirror pair.
Because the spot sizes at the two positions of 180~mm difference were measured to be the same,
a collimated condition was established.
The spot size injected to the irradiation system
was estimated as 16.5~mm (horizontal) $\times$ 10.5~mm (vertical) in RMS size of the photon density distribution.

%%%%%%%%%%%%%%%%%
%%%%%%%%%%%%%%%%%

\begin{figure}[htb]
	\begin{center}
 	 \includegraphics[bb=0 0 410 320, width= 0.8\linewidth]{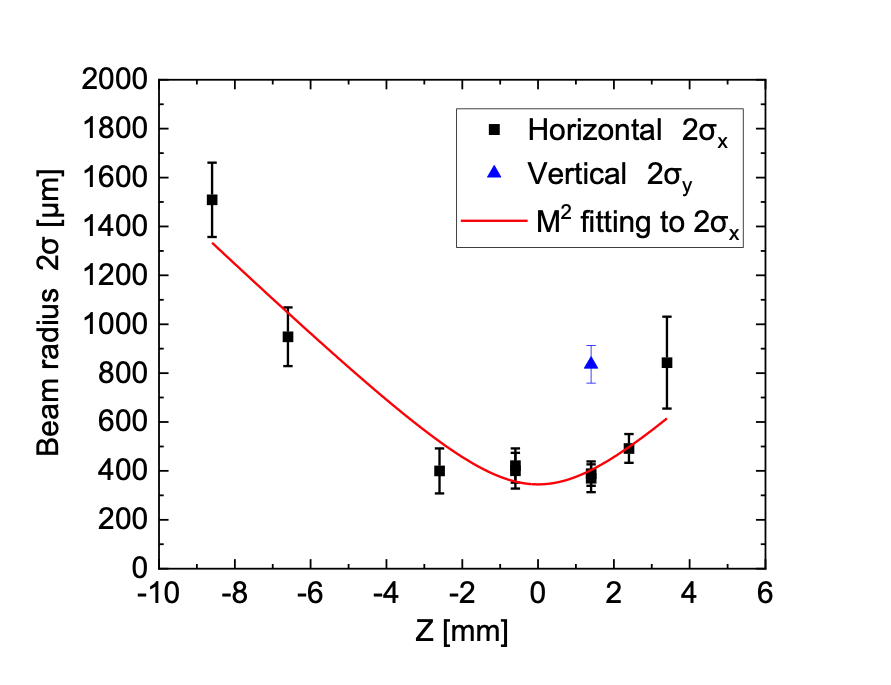}
	  \caption{
	  Spot size on the sample in the irradiation setup.
	  The measurement was performed with a knife-edge scan.
	  The data were fitted with a hyperbolic function to estimate the minimum size and 
	  beam quality ($M^2$). 
}
	\label{fig:knifeedgescanresult}
\end{center}
\end{figure}

In the irradiation system,
the spot size on the sample was measured by knife-edge scans.
By scanning the knife edge on the radiation spot with a 200~$\mu$m step
while measuring the transmission energy,
the spot size can be estimated.
Figure \ref{fig:knifeedgescanresult} shows the estimated spot size 
as a function of the longitudinal position of the sample stage.
The beam radius (2$\sigma$) are  plotted with the longitudinal position ($z$)
and fitted by an analytical Gaussian beam propagation equation to estimate beam quality factor $M^2$
%\cite{skinner, bilger, iso_m2}. 
\cite{iso_m2}. 
The minimum spot on the sample was $0.345 \pm 0.040$  mm (2-RMS size, $2\sigma_x$)
 in $x$ direction.
The 2-RMS size for $y$ direction measured only one $z$ position.
The $M^2$ was estimated as $8.1\pm0.85$ for x direction.
The measured macro-pulse energy ($E$) was 207 nJ, 
due to clipping by the 25.4 mm aperture of the shutter and the parabola mirror. 
The peak fluence ($Fp =2E/(\pi (2\sigma_x) (2\sigma_y$)) is estimated as $5.5\times10^{-5}$~J/cm$^2$ (for a macro-pulse) 
and $0.42\times10^{-6}$~J/cm$^2$ (for a pulse)  
with the measured beam size of $2\sigma_x=385~\mu$m and $2\sigma_y=625~\mu$m. 
These information are indispensable for application experiments to quantitatively evaluate the effect of mid-infrared FEL irradiation 
on various material.

%
%%%%%%%%%%%%%%%%%
%%%%%%%%%%%%%%%%%

%=============================
\section{DISCUSSION}
\label{sec:discussion}
%=============================

A demonstration experiment of mid-infrared SASE FEL targeting a wavelength of 20~$\mu$m
was performed at the cERL.
In spite of the adverse circumstance
that the beam quality was strongly affected by space charge effects,
we developed a procedure to transport the beam
in a controlled manner to the exit of the FEL layout.
The procedure for FEL tuning
has been established.

The FEL output energy 
was estimated as 5.8~nJ/pulse at 20~$\mu$m wavelength
at the exit of the second undulator. 
Comparing the extracted FEL energy 
with the beam energy of the 60~pC bunch at 17.7~MeV, which is approximately 1~mJ,
the beam-to-radiation conversion efficiency is calculated as $6\times10^{-6}$.
The obtained efficiency is rather lower than 
that naively expected with the Pierce parameter.
One reason for the discrepancy is the slippage effect.
According to a realistic three-dimensional simulation,
the experimental results can be understood.
In fact, the measured FEL energy was higher than 
that from the naive simulation shown in Fig. \ref{fig:powersim}.
The measured FEL energy at the exit of the first undulator 
was 0.72~nJ/bunch.
It is about 10 times higher than the energy from the simulation.
There appears to be an effect that enhances the start-up of the FEL process.
We guess that
there existed a fine structure that cannot be reproduced in the simulation input
in the initial bunch density distribution.
The coherent spontaneous emission (CSE) from 
the fine structure worked as a seed.
The kinks that appeared in the longitudinal phase space distribution due to microbunching instability (MBI)
as shown in Fig. \ref{fig:simslice}
might have worked as the seed of the CSE.
In the simulation,
we can reproduce the effect 
by introducing an empirical parameter of 0.02\% initial energy modulation.

The power gain in the second undulator with respect to the first undulator output
was measured in the range of 5 to 10 after the final tuning.
This is evidence of the contribution of the FEL to the amplification effect,
which is not due to just the CSE.
The amplification factor is consistent with the expectation from the simulation
in Fig. \ref{fig:powersim}.
The strong dependence of the FEL output on the incoming beam position
shown in Fig. \ref{fig:steeringresponse}
is another piece of evidence.

We performed beam tuning for FEL optimization approximately five times
after the initial machine parameters in the operation period.
The obtained FEL output power varied by approximately 20\% in each trial on different days.
In addition, the optimum $R_{56}$ of the arc slightly changed 
as can be observed in Fig. \ref{fig:r56response}.
The repeatability of the machine was not too bad as the first commissioning.

In the burst mode beam operation,
the beam energy changes in the macro-pulse
due to the beam loading effect.
The effect was estimated as 0.5\% of the beam energy shift.
Considering the resonance condition of the undulator,
it changes the radiation wavelength by approximately 1\%.
It is still much smaller than the relatively wide spectrum width shown in Fig. \ref{fig:simspectrum}.
However, because the FEL beam tuning is sensitive to the beam condition,
the slight energy difference might affect the optimum condition of each bunch.
To check the effect,
we changed the macro-pulse width to 50\%, that is from 130 bunches to 65 bunches.
The FEL power then decreased to approximately 50\%.
This observation is evidence that most of the bunches in the macro-pulse
contribute to the FEL output.

The undulator tapering scheme was applied to counteract the energy chirp in the bunch
and enhance the FEL power.
When we considered the undulator tapering in the design phase,
we repeated the simulation, changing the tapering coefficient of the undulators.
With the expected beam distribution used at the time,
the simulation showed the FEL power enhancement 
by a factor of 5 as a result of setting the tapering coefficient to approximately $-0.05$.
However, with the updated initial beam parameters in the actual beam operation,
the response to the tapering has changed.
To confirm whether undulator tapering worked,
a reliable comparison experiment
with and without tapering is necessary.
Although it will be the subject of one of our forthcoming studies,
it is outside the scope of this paper, which is about the first commissioning of the FEL.

The nominal FEL optimization was performed without using the chicane magnet.
We tried turning on the chicane after the nominal FEL optimization.
The FEL output showed some response to the current of the chicane magnet, i.e., the bump height.
The enhancement effect due to the chicane is still under study
and is outside the scope of this paper.

The radiation spectrum should be important for application experiments.
Because of the limited beam time,
we could not perform a detailed spectrum measurement.
However, from the simple measurements of the ZnSe plate transmission 
in Fig. \ref{fig:znsetransmission},
we can state that the center wavelength is consistent with the expectation,
and it can be adjusted by sliding the undulator.
For a detailed study,
we plan to install a spectrometer system in the next beam operation.

Beam profile measurement was carried out with the setup supposed for irradiation experiments. 
The knife edge measurement gives information on the focusing properties of FEL and the peak energy fluence on the sample. 
The estimated energy fluence ($\sim 5.5\times 10^{-5}$ J/cm$^2$ for one macro pulse) was smaller 
than the threshold fluence to achieve modification of resins \cite{sato}.  
Increase in the fluence by increasing the output energy and improvement of the beam quality 
together with tight focusing setup will realize various application experiments 
(laser processing, nonlinear phenomena) in near future.

To the best of our knowledge,
this is the first experiment 
on an SASE FEL installed in an ERL-type accelerator layout.
The most important advantage of the ERL layout
is the possibility of handling a high average current beam.
To investigate this possibility,
the following studies will be necessary:
Transport the beam to the main linac cavity in the deceleration phase
and safely guide it to the beam dump.
For that, 
the efficiency of energy recovery should be optimized
and the energy spread has to be managed in the momentum aperture of the beam dump line.
The beam loss must be reduced to the allowable level 
determined by the shielding wall of the accelerator.
The beam loss at the FEL hole mirrors
and the narrow chamber of the undulator must be studied.
We plan to extend the study
to cover high-current operation in energy recovery mode.
However, a discussion of this  is outside the scope of this paper.

%=============================
\section{CONCLUSION}
\label{sec:conclusion}
%=============================

The FEL is one of the promising light sources in the mid-infrared range with wavelength tunability. 
Due to the slippage effects,
the SASE FEL with a long undulator is not usually employed in the long wavelength range.
On the other hand,
the SASE FEL may have some advantages
such as the possibility of single-pulse operation
and the simplicity in establishing operating conditions,
and most importantly,
 it is scalabe to a future short-wavelength FEL.

We have constructed a mid-infrared SASE FEL setup
in the layout of the cERL, 
which is a test accelerator of an ERL scheme.
In spite of adverse circumstance of space charge effects 
due to the relatively low beam energy
and long transport line,
we have established a beam tuning procedure
to transport the beam through the undulator
and observed FEL emission at a wavelength of 20~$\mu$m.
A demonstration experiment of the FEL irradiation system
was conducted successfully and confirmed the stable operation of the machine.

The results show that
the layout of the cERL, a compact facility of the superconducting linac,
has potential for use as a mid-infrared light source.

%==============================
\section*{Acknowledgments}
%==============================

We appreciate the cERL members for their support of the beam operation. 
This paper is based on results obtained from a NEDO
Project Development of advanced laser processing with intelligence based
on high-brightness and high-efficiency laser technologies (TACMI project)
This work was partially supported by JSPS KAKENHI Grant Numbers 16H05991, 18H03473,
and 15K04747.
Construction of the cERL was supported 
by Photon and Quantum Basic Research Coordinated Development Program from the Ministry of Education, Culture, Sports, Science and Technology, Japan and by a Government (MEXT) Subsidy for Strengthening Nuclear Security.

%==========================
\appendix
%==========================
%---------------------------------------------------------------------------
\section{Space Charge Effects}
\label{sec:sc}
%---------------------------------------------------------------------------

The effect of the transverse space charge is not negligible
at 17.5~MeV with a bunch charge of 60~pC.
In a simple model of envelope equation for a DC beam approximation
\cite{sc_sarafini, sc_anderson},
the beam size development can be formalized as follows;
\begin{equation}
\sigma_x'' + K_x \sigma_x = 
\frac{I}{I_0 (\beta \gamma)^3 (\sigma_x + \sigma_y) }
+ \frac{\epsilon_{x}^2}{(\beta \gamma)^2 \sigma_x^3}
\end{equation}
and a similar equation can be formulated for the $y$ direction.
$K_x$ denotes the external focusing by the magnets,
and $\epsilon_{x}$ is the normalized emittance.
$\sigma_x$ and $\sigma_y$ are the horizontal and vertical beam sizes.
$I$ is the beam current, 
and $I_0$($\sim$17~kA) is a constant.
$\beta$ is the normalized velocity,
and $\gamma$ is the Lorentz factor.
From this equation, 
the defocusing force of the space charge 
can be rewritten as
\begin{equation}
K_{sc,x} = -\frac{I}{I_0 (\beta \gamma)^3 (\sigma_x + \sigma_y) \sigma_x}
\quad.
\end{equation}
Comparing the integration of this effect distributed along a typical length of the beam line
to the external focusing,
we can understand that
the effect is still small and can be treated as a perturbation
for understanding the overall beam optics of our case.

\begin{figure}[htb]
	\begin{center}
 	 \includegraphics[bb=0 0 550 470, width= 0.8\linewidth]{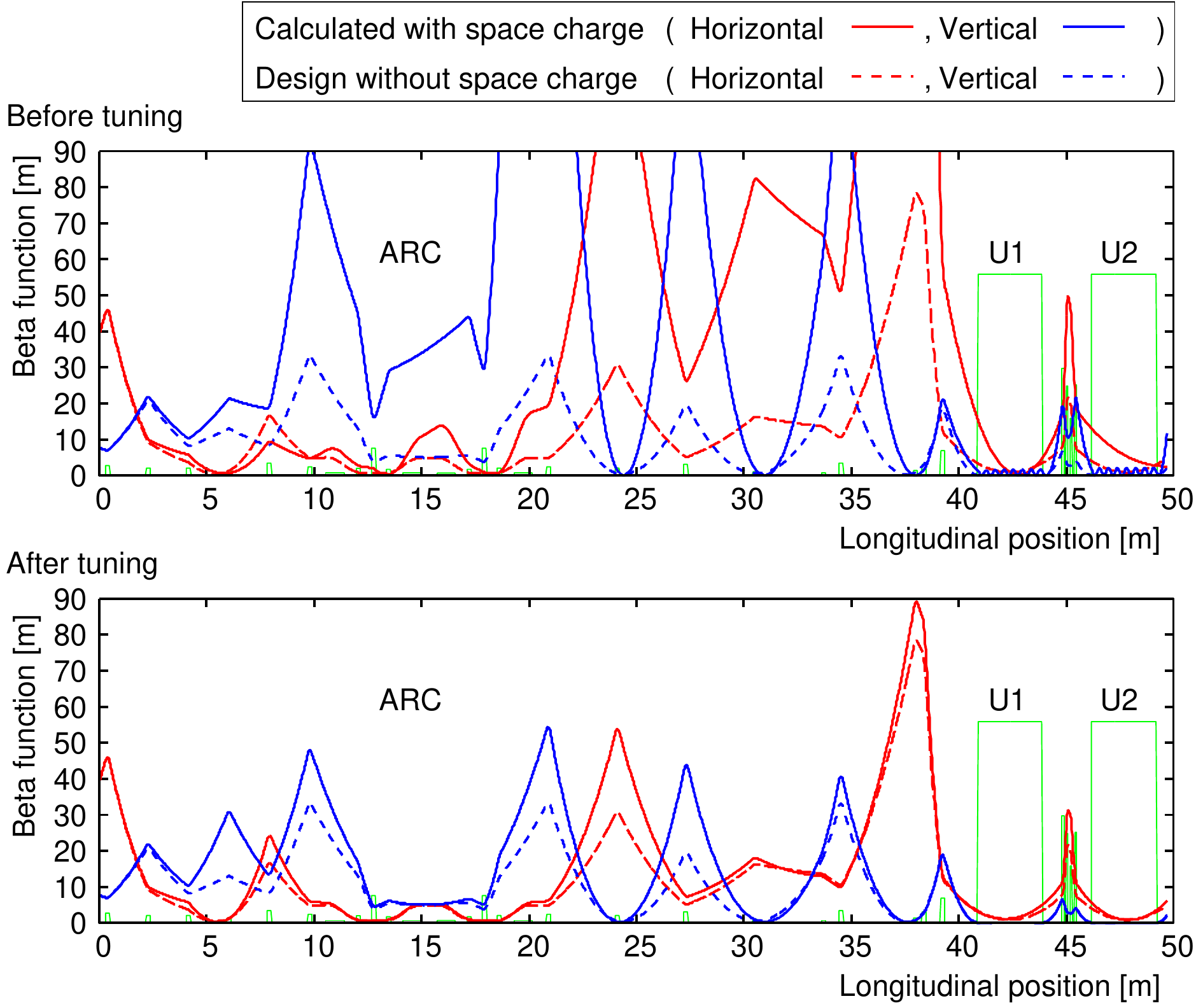}
	  \caption{
	  Beam optics (beta function) from the exit of the main linac to the end of the FEL section  
	  calculated based on the envelope equation, including the transverse space charge effect.
	  In the case of the magnet settings designed ignoring the space charge,
	  the beta function deviates significantly from the design (top).
	  After applying a local correction from upstream,
	  the overall shape of the beta function agrees fairly well with the design (bottom).
}
	\label{fig:envelopesim}
\end{center}
\end{figure}

Figure \ref{fig:envelopesim} shows the calculated beam optics
including the transverse space charge effect produced by the envelope equation
and its comparison with the designed linear optics.
Although the optics can seriously deviate from the design without any correction,
it can be matched fairly well by interleaving matching tuning with an appropriate distance.
Note that this model treats the space charge effect only linearly,
and hence the emittance is conserved.
However, nonlinear effect can dilute the beam emittance.

%%%%%%%%%%%%%%%%%%%%%%%
%%%%%%%%%%%%%%%%%%%%%%%

\begin{figure}[htb]
	\begin{center}
 	 \includegraphics[bb=0 0 600 400, width= 0.8\linewidth]{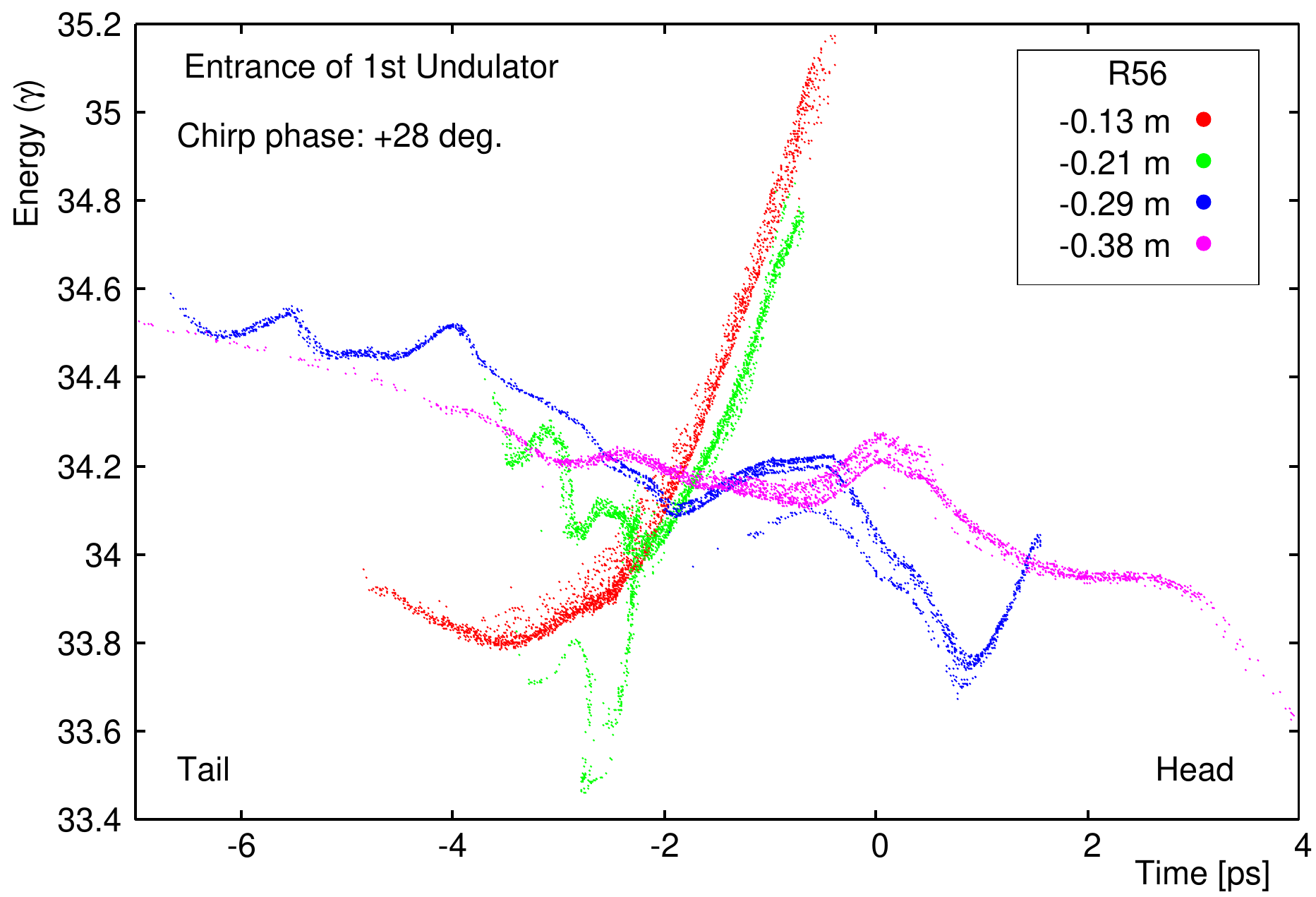}
	  \caption{
	  Simulated longitudinal phase space distribution at the entrance of the FEL section.
	  The distribution depends strongly on the bunch compression condition.
	  Corresponding to the experimental condition,
	  the chirp phase was set to be $+28^\circ$.
	  Four cases of $R_{56}$ are shown; $R_{56}=-0.29$~m is the experimental case.
	  Due to the LSC effect, the complicated shape appears
	  especially in the overbunching cases.
}
	\label{fig:lscdistribution}
\end{center}
\end{figure}

Due to the longitudinal space charge (LSC) effect,
the energy spread of the bunch increases in beam transport; i.e.,
the head part of the bunch is accelerated, and the tail part is decelerated.
The combination of  the LSC effect with the bunch compression in the arc section
severely disturbs the distribution in the longitudinal phase space \cite{lsc_huang}.
Figure \ref{fig:lscdistribution} shows 
examples of the phase space distribution at the entrance of the FEL undulator
calculated by elegant tracking.
A complicated oscillation structure is formed in the arc section.
This effect is called microbunching instability (MBI).

%%%%%%%%%%%%%%%%%%%%%%%
%%%%%%%%%%%%%%%%%%%%%%%

%---------------------------------------------
\section{Undulator Tapering}
%---------------------------------------------

Based on the beam transport simulations,
we found that
due to the space charge effect,
the energy spread of the bunch at the entrance of the undulator became large,
and it has a longitudinal energy chirp.
To achieve an FEL gain throughout the undulator,
the resonance condition of the FEL needs to be retained
during the slippage over the full bunch length. 
To counteract the shift of the resonance condition due to the energy chirp,
we utilized the undulator tapering scheme
\cite{sparc}.

From the resonance condition of the undulator
\begin{equation}
\lambda = \frac{\lambda_u}{2 \gamma^2} (1 + a_w^2) \quad,
\end{equation}
and considering the slice energy change $\Delta \gamma$ and the  undulator strength change $\Delta a_w$
in one period of the undulator,
the following relation is needed to keep the resonance condition in slip:
\begin{equation}
\frac{\partial \lambda}{\partial a_w} \Delta a_w = \frac{\partial \lambda}{\partial \gamma} \Delta \gamma
\quad .
\end{equation}
This results in the following relation:
\begin{equation}
\frac{\Delta a_w}{\Delta \gamma} = \frac{1+a_w^2}{\gamma a_w}
\quad .
\end{equation}
The undulator tapering should be designed according to this relation.

\begin{figure}[htb]
	\begin{center}
 	 \includegraphics[bb=0 0 600 400, width= 0.8\linewidth]{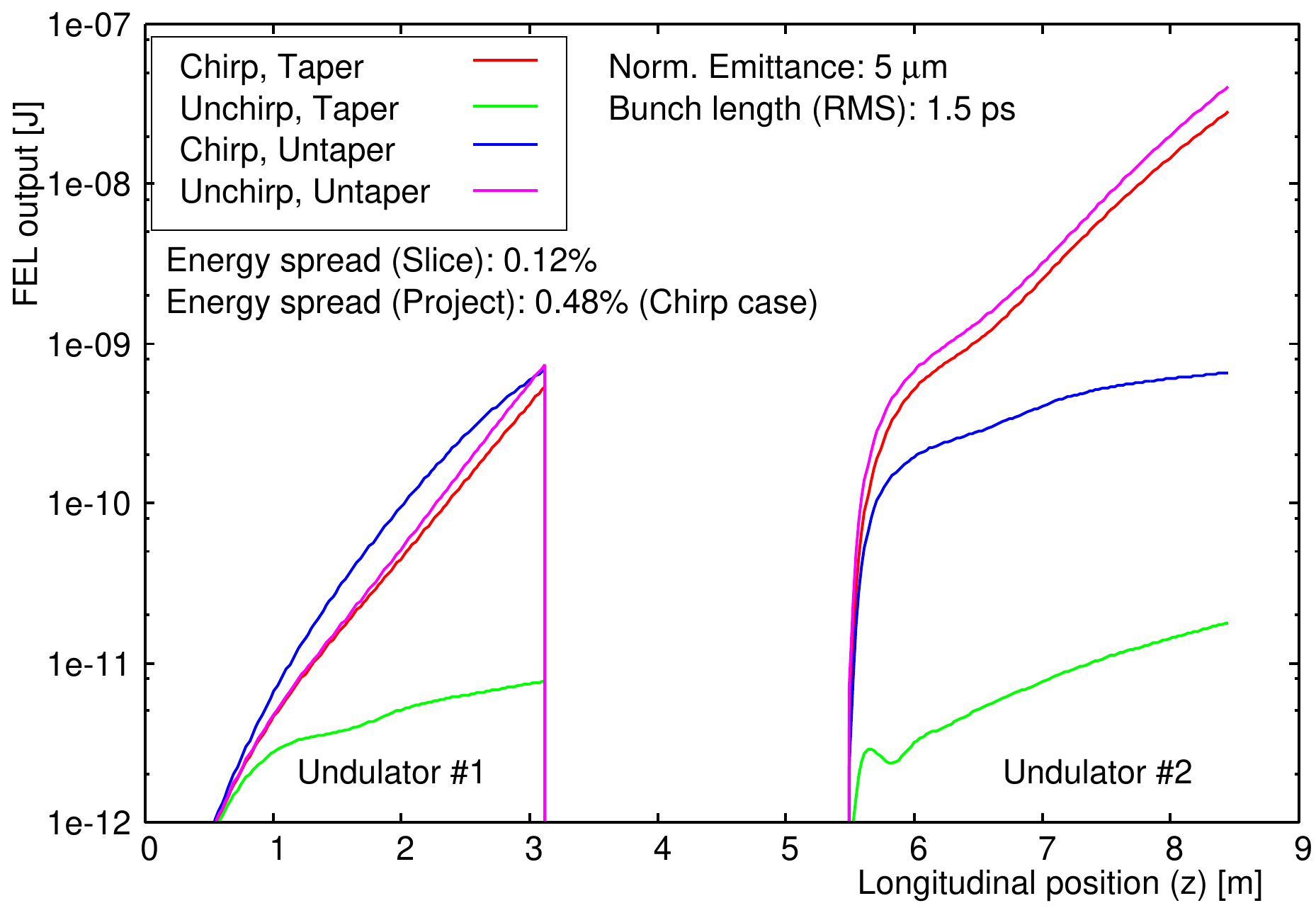}
	  \caption{
	  Example simulation showing the effect of undulator tapering.
	  Assuming a Gaussian distributed linearly chirped beam,
	  development of FEL power along the undulator was calculated.
	  Compared with the ideal case, unchirped beam with an untapered undulator,
	  the power was limited in the chirped and untapered cases.
	  The output power could be recovered to the ideal case 
	  by using tapered undulator.
}
	\label{fig:chirpandtaper_calc}
\end{center}
\end{figure}

Figure \ref{fig:chirpandtaper_calc} shows a simulation example
demonstrating the effect of the tapering scheme.
Even with a chirped beam,
the FEL power can be recovered with an optimized tapered undulator
to almost the same level
as in the ideal case of an unchirped beam with a non-tapered undulator.

%%%%%%%%%%%%%%%%%%%%%%
%%%%%%%%%%%%%%%%%%%%%%

%==============================
\section{Orthogonal Combined Knob }
\label{sec:knobs}
%==============================

To match the beam optics,
namely four Twiss parameters,
to the design at a position,
at least four quadrupole magnets are necessary.
The matching tuning is difficult and time consuming,
involving repeating manual trials using one magnet at each step.
To improve the efficiency of the beam tuning,
we prepared orthogonal knobs by combined actions using four magnets simultaneously.

To define orthogonality,
we used a complex representation of beam parameters.
This is by analogy with the technique used in  Gaussian beam optics of the laser beam \cite{yariv},
which has exactly the same mathematical structure.
Using the $\alpha$ and $\beta$ of the Twiss parameters at the target position,
the beam is represented as a point in the upper half area of the complex plane as
\begin{equation}
q = -\frac{\alpha \beta }{1+\alpha^2} + \frac{\beta}{1+\alpha^2}i \quad .
\end{equation}
We consider combined knobs
that move the beam to the horizontal (real axis) or vertical (imaginary axis) direction
in the plane. 
In the sense of two orthogonal directions, we call the knobs orthogonal.
The real and imaginary motions can be understood as 
shifting the longitudinal position of the beam waist while maintaining the size of the waist
and
changing the size of the beam waist while maintaining the longitudinal position of the waist,
respectively.
Note that because
the beam line components are represented as linear fractional transformations
that preserve angles,
the orthogonality is well defined in beam transport.
The orthogonal knobs were used in various cases in the beam tuning.

\bibliography{felbibfile}

\end{document}